# Title: Significantly Enhanced Vacancy Diffusion in Mn-containing Alloys

**Authors:** Huaqing Guan[1,2], Hanwen Cui[1,2], Ning Ding[3], Kuo Yang[4,5], Siqi Jiang[1,2], Yanfei Sui[1,2], Yuanyuan Wang[1,2], Fuyang Tian[3*], Zhe Li[4,5*], Shuai Wang[6*], Pengfei Zheng[7], Chenyang Lu[8], Qiu Xu[9], Levente Vitos[10], Shaosong Huang[1,2*]

**Affiliations:**

[1]School of Physics, Dalian University of Technology, Dalian, 116024, China.

[2]Key Laboratory of Materials Modification by Laser, Ion and Electron Beams (Dalian University of Technology), Ministry of Education, Dalian, 116024, China.

[3]Institute for Applied Physics, University of Science and Technology Beijing, Beijing, 100083, China.

[4]State Key Laboratory on Tunable laser Technology, Ministry of Industry and Information Technology Key Lab of Micro-Nano Optoelectronic Information System, School of Science, Harbin Institute of Technology (Shenzhen), Shenzhen, 518055, China

[5]Guangdong Provincial Key Laboratory of Semiconductor Optoelectronic Materials and Intelligent Photonic Systems, Harbin Institute of Technology (Shenzhen), Shenzhen 518055, China

[6]Department of Mechanical and Energy Engineering, Southern University of Science and Technology, Shenzhen, 518055, China.

[7]Southwestern Institute of Physics, Chengdu, 610000, China.

[8]School of Nuclear Science and Technology, Xi'an Jiaotong University, Xi'an, 710049, China.

[9]Institute for Integrated Radiation and Nuclear Science, Kyoto University, Osaka, 590-0494, Japan.

[10]Department of Materials Science and Engineering, Royal Institute of Technology (KTH) Brinellvägen 23, SE-10044 Stockholm, Sweden

*Corresponding authors. Email: huangss@dlut.edu.cn (S.H.); fuyang@ustb.edu.cn (F.T.); zhe.li@hit.edu.cn (Z.L.); wangs@sustech.edu.cn (S.W.)

**Abstract:** Manipulating point defects for tailored macroscopic properties remains a formidable challenge in materials science. This study demonstrates a proof-of-principle for a universal law involving element Mn, significantly enhancing vacancy diffusion through an unprecedented anomalous Friedel Oscillations phenomenon, across most metals in the periodic table. The correlation between Mn-induced point-defect dynamic changes and intrinsic macro-properties is robustly validated through the first-principles theory and well-designed experiments. The physical origin stems from Mn's exceptionally large effective intra-elemental *3d* electron interactions, surpassing the Coulomb attraction induced by vacancy and disrupting the electron screening effect. Given the ubiquitous nature of vacancies and their recognition as the most crucial defects influencing nearly all physical and mechanical properties of crystalline materials, this outcome may drive advances in a broad domain.

**Main Text:**

The utilization of alloys has been fundamental to human civilization for millennia, initially guided by empirical observations. A pivotal shift took place in the early 20th century with the recognition of defects as crucial factors in this domain. This understanding has propelled the development of superior materials across various applications. In materials science, unraveling the intricate relationship between defects and properties is essential for achieving desired material characteristics. Rooted in quantum mechanics (*1*) and band theory (*2*), particularly focusing on electronic behavior, this comprehension essentially influences alloy properties. While the electron theory of metals (*3*) has made substantial progress in developing new materials, it still faces limitations. For instance, the Hume-Rothery electron concentration rule struggles to apply when alloys contain transition metal (TM) elements as major constituents (*4*). The exploration of the phase stabilization mechanism for TM-bearing alloys remains unsettled in physical metallurgy (*5, 6*).

Concentrated Solid-Solution Alloys (CSAs) and High Entropy Alloys (HEAs), especially *3d*-transition metal HEAs like CrMnFeCoNi, disrupting traditional alloy design principles, garnering attention for their unconventional strengths, remarkable irradiation resistance, and high-temperature stability (*7-10*). These materials, characterized by high concentrations of multiple elements, offer an extensive compositional space. This presents both challenges and opportunities for advancing metallic electronic theory, extending it to these nonperiodic materials. In this context, we tried to explore how the chemical complexity specific to each element affects the evolution of point defects and the resultant macro-properties. Our demonstration highlights that the vacancy, as the most fundamental defect in alloys, exhibits significant tuning of diffusion from a metallic electron perspective. This phenomenon, unprecedented in alloys, establishes a correlation between dynamic alterations in point defects and metallic electron theory through quantum mechanical methods. The consequential changes in macroscopic properties are convincingly validated through well-designed experiments.

**Substantially reduced vacancy diffusion energy facilitated by Mn**

To address the challenges presented by intricate local atomic environments in CSAs, we employed a Density Functional Theory (DFT) approach, specifically the Similar Atomic Environment (SAE) method. This method transforms structural modeling into a minimization problem, optimizing configurations and enhancing the characterization of short-range order in solid solutions. Consequently, it offers a more thorough grasp of the connection between chemical electronic structures and defect evolution (*11*). The Climbing-Image Nudged Elastic Band (CI-NEB) method is utilized to determine vacancy diffusion energy. This involves removing one atom from the structure and calculating the migration barrier for all the nearest neighbor (1nn) atoms as they move into the created vacancy. For our study, supercells are constructed for three alloys ($Ni_{0.6}Co_{0.2}X_{0.2}$, where X is Co, Fe, and Mn). These supercells share the same size and atomic environments, differing only in the presence of the X solute atom (**Fig. 1A**). To isolate the impact of the local X element on vacancy diffusion, we intentionally selected vacancies at the same positions in all alloys.

**Figure 1B** outlines the distribution of energy barriers for the diffusion of the 1nn Ni and Co atoms exchanging with vacancy within the $Ni_{0.6}Co_{0.2}X_{0.2}$ alloys. Remarkably, all those atoms within the $Ni_{0.6}Co_{0.2}Mn_{0.2}$ alloy display strikingly lower diffusion energies, averaging an astonishing 0.35 eV less in comparison to the other two alloys. This finding indicates that the effect of Mn on vacancy diffusion differs substantially from that of Co and Fe. The diffusion energy, represents the energy barrier that must be surmounted to move from the initial point (IP) to the saddle point (SP) due to the potential difference. In **Fig. 1C**, the potential energy surface (PES) of the identical diffusion path in the alloys is presented, using the same reference standard. It is evident that the potential energy at the SP position in $Ni_{0.6}Co_{0.2}Mn_{0.2}$ is markedly lower than those in the other two alloys.

**Electronic structure variations during diffusion and Mn-induced anomalous Friedel Oscillations**

The diffusion process entails interactions between diffusing atom and both the 1nn atoms and the vacancy. Notably, the *d*-electrons are highly localized. Even at the SP (nearest position to 1nn atoms), the diffusing atom minimally interacts with them electronically (**fig. S1**). This underscores the pivotal role of the interaction between the diffusing atom (*D*) and vacancy (*V*), the potential caused by which can be expressed as:

$$V_{D-V}(r) = h\phi_{D-V}(r) = \frac{1}{2}h(V_D(r) + V_V(r)) \quad (1)$$

Where *h* represents the concentration-dependent factor, while $V_D(r)$ and $V_V(r)$ denote the potentials induced by the atom and vacancy, respectively. Given that the diffusing atoms are the same, the difference of $V_{(D-V)}(r)$ is primarily from $V_V(r)$, which can be determined from the charge change $\Delta\rho$:

$$V_V(r) = \int_0^\infty \frac{\Delta\rho(r)}{|r - r_0|} dr_0 \quad (2)$$

Typically, $\Delta\rho$ exhibits a swiftly diminishing oscillatory behavior as the distance from the vacancy expands. This phenomenon is commonly known as Friedel Oscillations (FO) *(12)*, originally proposed in 1950s and recently experimentally observed on an iron surface *(13)*. According to this theory, $\Delta\rho$ has the standard form as:

$$\Delta\rho(r) = A \frac{\cos(2k_F r + \varphi)}{r^3} \quad (3)$$

where both the amplitude *A* and the phase $\varphi$ are dependent on the vacancy potential and the host metal. **Figure 2A** presents a comparison between the actual changes and the fitting results by FO theory in electron charge density along <001> directions. Three-dimensional visualizations of charge density change, using a consistent isosurface value of 0.08 e/Å, are also displayed in **Fig. 2B**. In $Ni_{0.6}Co_{0.2}Co_{0.2}$ and $Ni_{0.6}Co_{0.2}Fe_{0.2}$ alloys, the formation of vacancies results in a conventional FO pattern of the charge density, which is confined to the 1nn atoms. Conversely in $Ni_{0.6}Co_{0.2}Mn_{0.2}$, the oscillations extend beyond the 1nn atoms, reaching distances of more than 4nn, with no directional variability, revealing a novel phenomenon as anomalous FO.

**Physical origin of Mn's effects**

The observed phenomenon in $Ni_{0.6}Co_{0.2}Mn_{0.2}$ underscores the inapplicability of the classical FO formula. In the derivation of **Eq. (3)**, Friedel primarily addressed the issue of electron disturbances arising from the presence of solutes or defects in dilute solid-solution alloys *(12)*. However, in this study, our focus lies on exploring the effects of alloying elements on vacancies in CSAs. It is crucial to take into account factors such as atomic size mismatch, point charge approximation due to vacancies, and scattering phase effects resulting from alloying elements other than the generated vacancies (*14*). Our system involves *3d* transition metals with a radius difference below 5% (**table S1**). The point charge values by DFT within generated vacancy in $Ni_{0.6}Co_{0.2}Mn_{0.2}$ is 0.25, which is inconsequential compared to the charge density change within the 1nn regions ($\Delta\rho_{1nn}$) 6.71 (**fig. S2** and **fig. S3**). Thus, the scattering phase shift emerges as the most influential factor. We refit the charge density variation along the <001> direction in $Ni_{0.6}Co_{0.2}Mn_{0.2}$ as shown in **Fig.2C** by introducing the scattering phase shift term from Mn atoms, based on free-electron Green's function. The results exhibit a substantial alignment with the actual one by DFT, confirming the undeniable influence of Mn's scattering roles on electronic behavior.

The physical origin of these scattering effects can be explained as follows. **Figure 3** illustrates the complete process involving Mn, primarily driven by electronic interactions. The effects are intricately tied to a delicate balance between Mn's electronic hybridization with 1nn Ni and the effective Coulomb interaction among its intra-elemental *3d* electrons ($U_{eff}$). The crucial point is that the $U_{eff}$ of Mn is notably larger than that of Fe and Co, concurrently surpassing the Coulomb attraction induced by vacancies. The determination of $U_{eff}$ involves X-ray photoelectron spectroscopy (XPS), Auger Electron Spectroscopy (AES), and ultraviolet photo-electron spectroscopy (UPS) experiments, focusing on the LMM process (**fig. S5** and **fig. S6**) (*15, 16*). The experimental determination of $U_{eff}$ for X elements in $Ni_{0.6}Co_{0.2}X_{0.2}$ alloys is elaborated in **Table S2**. Conspicuously, $U_{eff}$ for Mn surpasses 4.6 eV, much larger than the approximately 2 eV values observed for Fe and Co. Pre-vacancy, the Mn DOS exhibits Lorentzian-type virtual bound states (VBS) above the host *d* bands. This VBS acts as an electron conduit, fostering hybrid coupling with the adjacent Ni atom (**Fig. 3A**) (*17*). The DOS overlap between Mn and neighboring Ni, as well as the heightened likelihood of Mn electrons appearing towards the vacancy direction in the local Electron Localization Function (ELF) plot, both signify the hybridization. Analysis by the Crystal Orbital Hamiltonian Population (COHP) method reveals that approximately two electrons in Mn participate in the hybridization with Ni atom (**fig. S7**). This robust hybridization is due to Mn's unique ground state of $d^5(^6S)$, which features broader energy levels (*18*), enhancing hybridization effects, as demonstrated by angle resolved photoemission and inverse photoemission experiments (*19*). According to Slater theory (*20, 21*), this hybridization competes with $U_{eff}$, leading to a reduction in intra-elemental electronic interactions. Post-vacancy formation, the cessation of hybridization triggers the restoration of correlation effects among Mn electrons. As depicted in **Fig. 3B**, splitting of the majority-spin DOS in Mn, along with the change in the probability of electron occurrence in Mn suggest a local moment self-consistent renormalization (*22-24*). Remember that $U_{eff}$ for Mn exceeds 4.6 eV, whereas the Coulomb attraction induced by vacancy generation is estimated at 2 eV, markedly smaller than the restored $U_{eff}$ for Mn, while larger than the $U_{eff}$ for Fe and Co (**table S2**). This discrepancy for Mn atom induces charge depletion towards the vacancy direction, as demonstrated in **Fig. 3C** (right) through the local deformation charge density. The localized reduction in charge compensation for the vacancy disrupts the electron screening effect, resulting in a decrease in potential energy between the vacancy and its 1nn atoms in the $Ni_{0.6}Co_{0.2}Mn_{0.2}$ (**Fig. 3D**). This condition facilitates the exchange of 1nn atoms with the vacancy, ultimately promoting an accelerated diffusion of vacancies.

**Coherence between Mn-induced alterations in point-defect dynamics and inherent macroscopic diffusion properties**

To validate the practical applicability of this theoretical process, temperature is an inevitable factor. The impact of temperature on the enthalpy for diffusion is, at most, on the order of $K_BT$ (*25*). Even at 1273 K, the value remains prominently modest, around 0.1097 eV. This is considerably less than the average reduced vacancy diffusion barrier in $Ni_{0.6}Co_{0.2}Mn_{0.2}$, by a factor of nearly 10.

Experimentally demonstrating the specific influence of Mn on atomic diffusion poses a significant challenge. Techniques like low-temperature in-situ positron tests and ultra-high-voltage electron microscopy enable indirect measurement of point vacancy dynamics (*26, 27*). However, they face limitations in establishing element-specific distinctions. The observed vacancy dynamics provide an average result for all elements, making it challenging to discern whether variations arise from the atoms themselves or their influence on other atoms. Given the substantial alterations in vacancy diffusion properties attributed to Mn, as evidenced by DFT calculations demonstrating minimal temperature dependence, we endeavor to validate this outcome through carefully designed reverse experiments. Specifically, we plan to macroscopically examine the diffusion coefficients of the three alloys at various temperatures and subsequently infer the origins of their microscopic mechanisms.

Three diffusion couples $Ni_{0.8}X_{0.2}/Ni_{0.6}Co_{0.2}X_{0.2}$, with X represents Co, Fe, and Mn, respectively, were prepared and annealed at high temperatures for 100 h and to investigate the X element effects on the diffusivity of Ni and Co, with only Ni and Co having concentration gradient (**fig. S8** and **fig. S9**). **Figure. 4A** shows the experimental measured concentration profiles of diffusion couples annealed at 900 °C, 1000 °C, and 1100 °C for 100 h by Electron Probe Micro-Analyzer (EPMA). By employing Boltzmann distribution fitting in Fick's second law, the diffusion coefficient ($D$) was calculated by measured data at various temperatures (**fig. S10** and **table S3**). The diffusion coefficient in Mn-containing diffusion couple is two orders of magnitude higher than those in the other two. In this vacancy-type diffusion process, with the similar vacancy equilibrium concentration (**fig. S11**) and identical concentration gradients of Ni and Co on both sides of the three diffusion couples, the presence of Mn conclusively demonstrates a significant enhancement in macro-scale vacancy diffusion.

The Arrhenius equation establishes the connection between macroscopic diffusion coefficient and microscopic diffusion mechanism. The logarithm of $D$ was graphed as a linear function of l/T, unveiling the diffusion activation energy ($Q$) through the slope (**Fig. 4B**, **Fig. 4D**, and **table S4**). In this case, $Q$ is the sum of the formation energy ($E_f$) and migration energy ($E_d$) of a vacancy. $E_f$ can be determined through electrical resistivity measurements at various temperatures. According to Mott and Jones theory (*28, 29*), the electrical resistance is anticipated to show a quadratic dependence at low temperatures. However, at higher temperatures, the measured resistance values surpass the extrapolated quadratic function. This extra resistance, $\Delta R$, is attributed to the presence of vacancies in thermodynamic equilibrium (**fig. S11**). The determined values of $E_f$ of the alloys by electrical resistivity measurements from 2-800 K are presented in **Fig. 4D**, indicating minimal variation among them. This implies that the key factor influencing differences in macro-level vacancy diffusion lies primarily in $E_d$. Evidently, take Co as an example, experimental findings disclose a vacancy diffusion barrier of 0.95 eV in $Ni_{0.6}Co_{0.2}Mn_{0.2}$, markedly lower than the values of 1.45 eV in $Ni_{0.6}Co_{0.2}Co_{0.2}$ and 1.52 eV in $Ni_{0.6}Co_{0.2}Fe_{0.2}$ (**table S5**). These values are very consistent with our DFT calculation results, which strongly prove that the influence of Mn on the microscopic diffusion barrier of vacancies can be extended to the macroscopic diffusion performance.

**Universality of Mn's effect**

**Figure 5** shows the deformation charge density and average vacancy diffusion barrier in nine commonly used metal pure systems and alloys featuring 20 at.% Mn through the introduction of a vacancy. In contrast to pure systems where only the charge density of 1nn atoms is impacted, alloys with Mn showcase extensive charge changes. Distinctively, the vacancy diffusion barrier in Mn-containing alloys is markedly lower than in their pure alloy counterparts. This reduction holds true across both transition and non-transition metals, consistently surpassing 0.2 eV. This observation underscores Mn's universal influence in facilitating vacancy diffusion, irrespective of alloy systems and structures.

**Conclusions**

Starting from the manipulation of atomic structures to achieve macroscopic targeted properties is almost the ultimate dream of material scientists. This study, spanning from theory to practical validation, establishes that the widespread and fundamental point defect—vacancy in materials can attain desired macroscopic performance through the aforementioned approach. Specifically, leveraging its unique *3d* electronic structure, Mn disrupts the electron screening effect through unprecedented anomalous Friedel Oscillations, thereby reducing potential energy variations and expediting vacancy diffusion. This work not only reveals an interesting principle that advances our understanding of metallic electronic theory but also deepens our insight into the essence of atomic diffusion in alloys. Importantly, we introduce a method capable of modifying the diffusion properties of individual vacancies and consequently their macroscopic behaviors—a universal principle applicable across nearly all metal systems in the periodic table. This finding opens new avenues for designing advanced metallic materials by tailoring vacancy properties in diverse alloying systems. For example, improving vacancy mobility in alloys is pivotal for deploying advanced nuclear energy systems, addressing material radiation damage rooted in the superior mobility of self-interstitials over vacancies.

**References and Notes**

1. A. Messiah, *Quantum mechanics* (Publisher, 2014).

2. J. Singleton, *Band theory and electronic properties of solids* (Publisher, 2001).

3. A. A. Abrikosov, *Fundamentals of the Theory of Metals* (Publisher, 2017).

4. U. Mizutani, M. Inukai, H. Sato, E. S. Zijlstra, *Electron Theory of Complex Metallic Alloys Physical metallurgy* (Publisher, 2014).

5. D. Shechtman, Metallurgical Aspects of Quasicrystals. *Mater. Sci. Forum* **22**, 1-10 (1987).

6. A. P. Tsai, J. Q. Guo, E. Abe, H. Takakura, T. J. Sato, A stable binary quasicrystal. *Nature* **408**, 537-538 (2000).

7. C. Lu, L. Niu, N. Chen, K. Jin, T. Yang, P. Xiu, Y. Zhang, F. Gao, H. Bei, S. Shi, R. Mo, I.M. Robertson, W.J. Weber and L. Wang, Enhancing radiation tolerance by controlling defect mobility and migration pathways in multicomponent single phase alloys, *Nat. Commun.* **7**, 13546 (2016).

8. B. Gludovatz, A. Hohenwarter, D. Catoor, E. H. Chang, E. P. George, R. O. Ritchie, A fracture-resistant high-entropy alloy for cryogenic applications. *Science.* **345**, 1153-1158 (2014).

9. D. Liu, Q. Yu, S. Kabra, M. Jiang, P. F. Kreutzer, R. Zhang, M. Payne, F. Walsh, B. Gludovatz, M. Asta, A. M. Minor, E. P. George, R. O. Ritchie, Exceptional fracture toughness of CrCoNi-based medium- and high-entropy alloys at 20 kelvin. *Science* **378**, 978-983 (2022).

10. Z. Rao, P. Tung, R. Xie, Y. Wei, H. Zhang, A. Ferrari, T. P. C. Klaver, F. Kormann, P. T. Sukumar, A. K. Silva, Y. Chen, Z. Li, D. Ponge, J. Neugebauer, O. Gutfleisch, S. Bauer, D. Raabe, Machine learning–enabled high-entropy alloy discovery. *Science* **378**, 78-85 (2022).

11. Materials and methods are available as supplementary materials.

12. J. Friedel, Metallic alloys, *Il Nuovo Cimento* **7**, 287-311 (1958).

13. T. Mitsui, S. Sakai, S. Li, T. Ueno, T. Watanuki, Y. Kobayashi, R. Masuda, M. Seto, H. Akai, Magnetic Friedel Oscillation at the Fe (001) Surface: Direct Observation by Atomic-Layer-Resolved Synchrotron Radiation 57Fe Mössbauer Spectroscopy. *Phys. Rev. Lett.* **125**, 236806 (2020).

14. J. S. Galsin, *Impurity scattering in metallic alloys* (Publisher, 2012).

15. E. Antonides, E. C. Jose, G. A. Sewatzky, LMM Auger spectra of Cu, Zn, Ga, and Ge. I. Transition probabilities, term splittings, and effective Coulomb interaction. *Phys. Rev. B* **15**, 4 (1977).

16. D. V. D. Marel, G. A. Sawatzky, Auger line shape in alloys. *Phys. Rev. B* **29**, 3073 (1984).

17. G. T. de Laissardiere, D. N. Manh, L. Magaud, J. P. Julien, F. C. Lackmann, D. Mayou, Electronic structure and hybridization effects in Hume-Rothery alloys containing transition elements. *Phys. Rev. B* **52**, 11 (1995).

18. D. V. D. Marel, C. Westra, G. A. Sawatzky, Electronic structure of Mn impurities in noble metals. *Phys. Rev. B* **31**, 4 (1985).

19. O. Rader, W. Gudat, C. Carbone, E. Vescovo, S. Blügel, R. Kläsges, W. Eberhardt, M. Wuttig, J. Redinger, F. J. Himpsel, Electronic structure of two-dimensional magnetic alloys: (2×2) Mn on Cu(100) and Ni(100). *Phys. Rev. B* **55**, 5404 (1997).

20. J. C. Slater, The electronic structure of metals. *Rev. Mod. Phys.* **6**, 209 (1934).

21. P. W. Anderson, Local Moments and Localized States. *Science* **201**, 4353 (1978).

22. C. Berthier, M. Minier, Evidence of screening charge depression in the vicinity of chromium and manganese impurities in aluminium at low temperature. *J. Phys. F: Metal Phys.* **3**, 1268 (1973).

23. S. Mu, J. Yin, G. D. Samolyuk, S. Wimmer, Z. Pei, M. Eisenbach, S. Mankovsky, H. Ebert, G. M. Stocks, Hidden Mn magnetic-moment disorder and its influence on the physical properties of medium-entropy NiCoMn solid solution alloys. *Phys. Rev. Mater.* **3**, 014411 (2019).

24. Y. Zhang, Y. N. Osetsky, W.J. Weber, Tunable chemical disorder in concentrated alloys: defect physics and radiation performance. *Chem. Rev.* **122**, 1 (2021).

25. Z. Fan, V. Gihan, K. Jin, M. L. Crespillo, H. Bei, W. J. Weber, Y. Zhang, Temperature-dependent defect accumulation and evolution in Ni-irradiated NiFe concentrated solid-solution alloy. *J. Nucl. Mater.* **519**, 1-9 (2019).

26. H. Fujita, T. Tabata, K. Yoshida, N. Sumida, S. Katagiri, Some applications of an ultra-high voltage electron microscope on materials science. *J. Appl. Phys.* **11**, 1522 (1972).

27. Z. Balogh, G. Schmitz, *Diffusion in metals and alloys* (Publisher, 2014).

28. N. F. Mott, H. Jones, *The theory of the properties of metals and alloys* (Publisher, 1936).

29. C. J. Meechan, R. R. Eggleston, Formation energies of vacancies in copper and gold. *Acta metal.* **2**, 680-683 (1954).

30. G. Kresse, J. Hafner, Ab initio molecular-dynamics simulation of the liquid-metal-amorphous-semiconductor transition in germanium. *Phys. Rev. B* **49**, 14251 (1994).

31. G. Kresse, J. Furthmüller, Efficiency of ab-initio total energy calculations for metals and semiconductors using a plane-wave basis set. *Comput. Mater. Sci.* **6**, 15 (1996).

32. G. Henkelman, B. P. Uberuaga, H. Josson, A climbing image nudged elastic band method for finding saddle points and minimum energy paths. *J. Chem. Phys.* **113**, 22 (2000).

33. F. Tian, D. Y. Lin, X. Gao, Y. F. Zhao, H. F. Song, A structural modeling approach to solid solutions based on the similar atomic environment. *J. Chem. Phys.* **153**, 034101 (2020).

34. H. Guan, S. Huang, J. Ding, F. Tian, Q. Xu, J. Zhao, Chemical environment and magnetic moment effects on point defect formations in CoCrNi-based concentrated solid-solution alloys. *Acta Mater.* **187**, 122 (2020).

35. P. Bag, Y. C. Su, Y. K. Kuo, Y. C. Lai, S. K. Wu, Physical properties of face-centered cubic structured high-entropy alloys: Effects of NiCo, NiFe, and NiCoFe alloying with Mn, Cr, and Pd. *Phys. Rev. Mater.* **5**, 085003 (2021).

36. L. Pauling Atomic radii and interatomic distances in metals. *J. Amer. Chem. Soc.* **69**, 3 (1947).

37. P. Rennert, Influence of a resonance state on the density oscillations around an impurity. *Phys Status Solidi B* **50**, 1 (1972).

38. F. Mezei, A. Zawadowski, Kinematic Change in Conduction-Electron Density of States Due to Impurity Scattering. I. The Problem of a Single Impurity. *Phys. Rev B* **3**, 1 (1971).

39. D. V. D. Marel, G. A. Sawatzky, Direct Observation of the Exchange-Split Virtual Bound State in Dilute Mn Alloys. *Phys. Rev. Lett.* **53**, 2 (1984).

40. L. I. Yin, E. Yellin, I. Adler, X-Ray Excited LMM Auger Spectra of Copper, Nickel, and Iron. *J. Appl. Phys.* **42**, 3595 (1971).

41. G. A. Sawatzky, E. Antonide, The electronic structure and electron correlation effects studied by XPS. *J. Phys. Colloq.* **37**, c4 (1976).

42. D. V. D. Marel, G. A. Sawatzky, Electron-electron interaction and localization in d and f transition metals. *Phys. Rev. B* **37**, 10674 (1988).

43. S. P. Kowalczyk, R. A. Pollak, F. R. McFeely, L Ley, D. A. Shirley, LMM Auger Spectra of Metallic Copper and Zinc: Theory and Experiment. *Phys. Rev. B* **8**, 6 (1973).

**Acknowledgments:** We thank Dalian University of Technology Instrumental Analysis Center for supports in providing XPS, AES, UPS, EPMA, and resistant measurements. Additionally, we extend our appreciation Tongmin Wang and Huijun Kang for their assistance with high temperature electrical resistivity measurements (300–800 K) and Fengyun Yu for support with the Electron Probe Micro-Analyzer. The Beijing Paratera is gratefully acknowledged for providing computational resources with the high-performance computer for the DFT calculations in the present work.

**Funding:** The research was supported by the National Natural Science Foundation of China (12275045 to S.H.); NSFC Grant (No. 52371174 to F.T.); the National Natural Science Foundation of China (91961102 to Z.L.); the Shenzhen research funding (GXWD20231130102757002 to Z.L.); the Innovation Capability Support Program of Shaanxi Province (2023-CX-TD-49 to Z.L.); National Key R&D Program of China (No.2022YFB4600700 to S.W.); the Swedish Research Council for Sustainable Development (Project nr. 2023-00543_Formas to L.V.); the Swedish Foundation for Strategic Research (Project nr. SM23-0041 to L.V.), and the Carl Tryggers Foundation (Project nr. CTS 22:1970 to L.V.)

**Author contributions:** S.H. formulated the original idea. F.T. provided supercell structures for DFT calculations. H.G., H.C. and N.D. performed DFT calculations. S.H. designed the series experiments. H.G. and M.W. prepared the samples for experiments. Y.S. and M.W. conducted AES and XPS experiment. H.G. carried out the diffusion-couple experiments. K.Y. performed the electrical resistivity measurements under the supervision of Z.L. The calculation and experimental data were analyzed by H.G. under the supervision of S.H. Result analysis and discussions were actively participated in by Y.W., P.Z., S.W., C.L., X.Q., and L.V. The manuscript was drafted by H.G and S.H., with all authors contributing. All authors reviewed and commented on successive drafts of the manuscript.

**Competing interests:** Authors declare that they have no competing interests.

**Data and materials availability:** The data that support the findings of this study are available in the main text or the supplementary materials.

**Supplementary Materials**

Materials and Methods

Supplementary Text

Figs. S1 to S11

Tables S1 to S5

References (*30-43*)

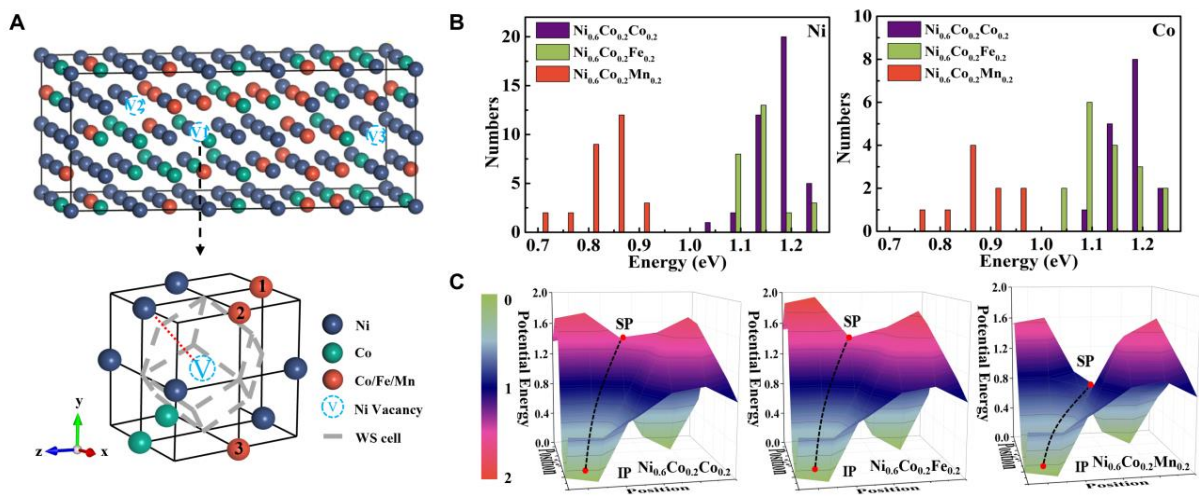

**Fig. 1 Atomic structure, vacancy diffusion energy, and potential energy surface in Ni$_{0.6}$Co$_{0.2}$X$_{0.2}$ (X=Co/Fe/Mn) alloys.** (**A**) Supercell structure of Ni$_{0.6}$Co$_{0.2}$X$_{0.2}$ (X=Co/Fe/Mn) by SAE method. (**B**) The distribution of energy barriers for the diffusion of 1nn atoms other than X exchanging with vacancy in the V1, V2, and V3 cases within the supercell shown in (A), Ni atoms (left) and Co atoms (right). (**C**) The potential energy surface (PES) along the identical diffusion path (illustrated by the red dotted line in (A)) in the alloys by using the same reference standard. The Ni atom diffuses from the initial point (IP) along the dotted line to the saddle point (SP).

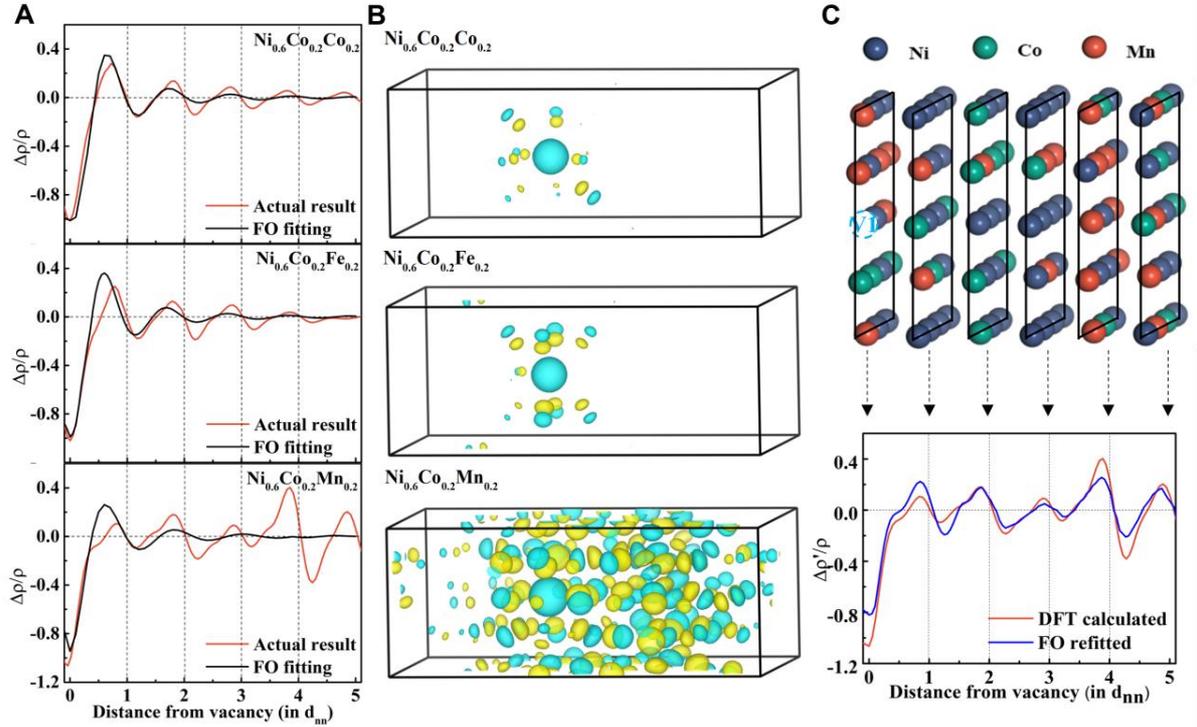

**Fig. 2 Charge distribution changes induced by vacancy formation in $Ni_{0.6}Co_{0.2}X_{0.2}$ (X=Co/Fe/Mn) alloys.** (**A**) Charge density changes along <001> direction with vacancy generation at the same position V1 in the three supercells. Distance from the vacancy is in scale of $d_{nn}$=2.45 Å. The actual (red line) and the fitting result (black line) from Friedel Oscillation theory are presented. (**B**) Three-dimensional deformation charge density after the vacancy generation, using the same isosurface of 0.08 e/Å. (**C**) The refitted curve by considering scattering phase shift of Mn atoms and the DFT calculated charge density variation along <001> direction in $Ni_{0.6}Co_{0.2}Mn_{0.2}$. The atomic environment of neighboring surfaces at varying distances from vacancy are delineated.

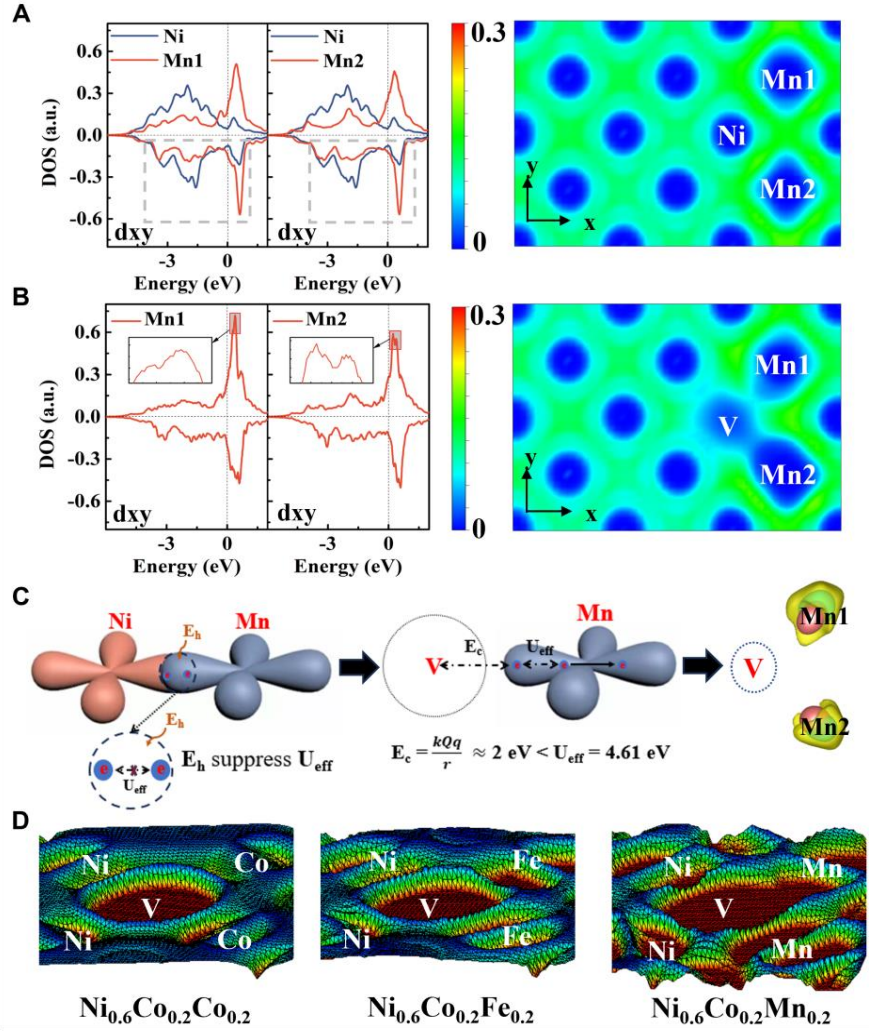

**Fig. 3 The physical origin of Mn's effects.** (**A**) Pre-vacancy, DOS of Ni atom that generate vacancy (V) and the nearest-neighbor Mn atoms (left), accompanied by local Electron localization function (ELF) (right), both along the <001> direction. The DOS overlap between Mn and neighboring Ni, as well as the heightened likelihood of Mn electrons appearing towards the vacancy direction in ELF plot, both signify the hybridization. (**B**) Post-vacancy formation, the DOS of Mn (left) and the ELF (right), both along the <001> direction. The splitting of the majority-spin DOS in Mn, along with the change in the probability of electron occurrence in Mn suggest a local moment self-consistent renormalization. (**C**) Schematic diagram of local electronic orbitals and electronic states changes of Mn, showing the balance between Mn's hybridization with nearest-neighbor Ni and the effective Coulomb interaction among Mn's intra-elemental *3d* electrons ($U_{eff}$), and its impact on the charge depletion around the created vacancy. **Left**: Pre-vacancy, hybridization between Mn and its adjacent Ni atom. **Middle**: Post-vacancy formation, the restored $U_{eff}$, surpassing the Coulomb attraction induced by the vacancy, leads to the displacement of electrons away from the vacancy site. **Right**: Charge depletion in between the created vacancy and nearest-neighbor Mn atoms along $d_{xy}$ direction. $E_h$ denotes hybridization effect, $E_c$ represents Coulomb energy induced by the generated vacancy. (**D**) The local energy landscape along [001] after vacancy formation.

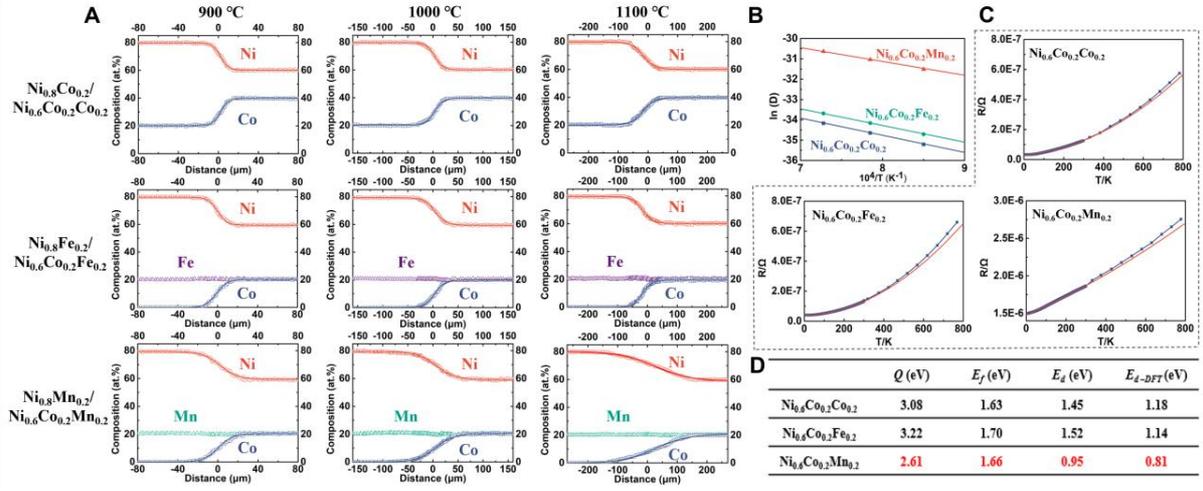

**Fig. 4 Diffusion couple and resistance measurement experiments.** (**A**) Concentration profiles of the three diffusion couples annealed at 900 °C, 1000 °C, and 1100 °C for 100 h, with experimental data (symbols) (Note that the X-axis scales vary across temperatures to effectively illustrate the distinct profile variations among different alloys at identical temperature). (**B**) Experimental obtained temperature (the inverse of the absolute value) dependence of the natural logarithm of diffusion coefficients. (**C**) Measured electrical resistivity of samples (dots), in the range of 2-800 K. The red curves are fitted from low-temperature (2-300 K) experimental data and extrapolated to high temperatures. (**D**) The vacancy diffusion activation energy ($Q$) for Co element and vacancy formation energy ($E_f$) obtained from diffusion couple experiment and resistance measurement, respectively; the diffusion energy ($E_d$) is obtained from the difference between the two and compared with the DFT calculation results ($E_{d-DFT}$).

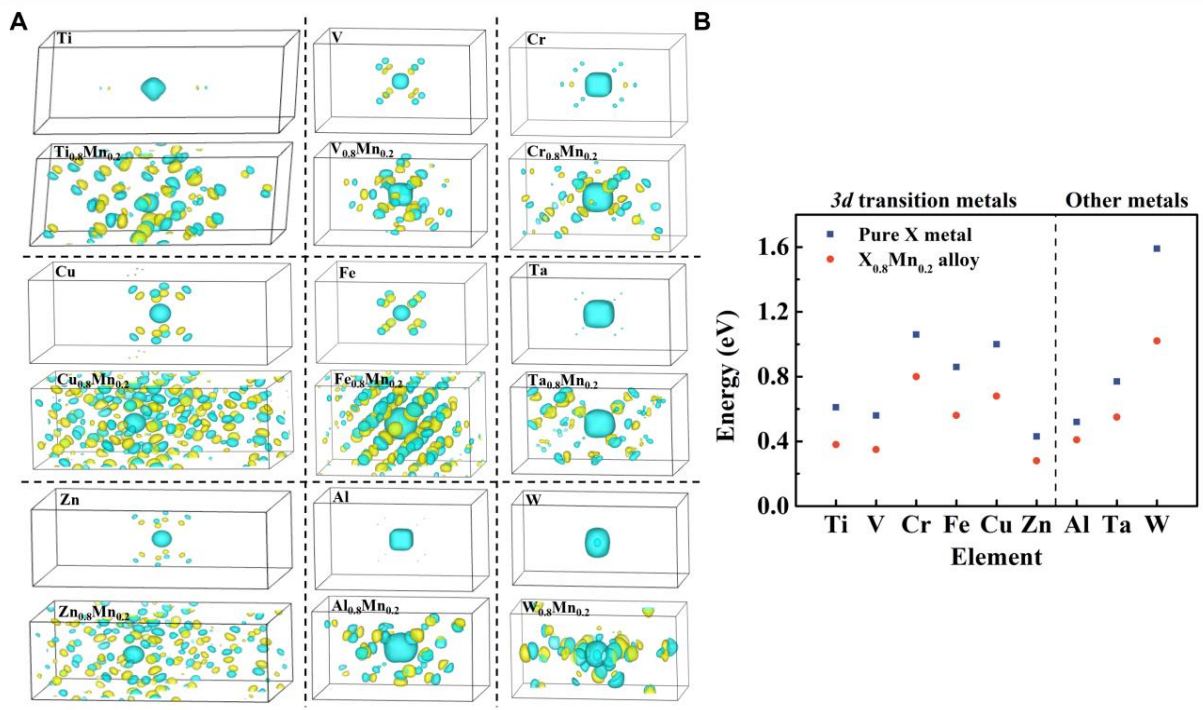

**Fig. 5 Charge distribution changes and average vacancy diffusion energies for a series of metals and their binary alloys with 20 at.% Mn content.** (**A**) Three-dimensional deformation charge density after a vacancy generation in Ti, V, Cr, Fe, Cu, Zn, Al, Ta, and W metals and their binary alloys with 20 at.% Mn content. (**B**) Average diffusion energy barrier of 1nn atoms exchanging with the generated vacancy shown in (A). Note the DFT+U with $U_{eff}$= 2 eV is used to accurately describe electron correlations in $Fe_{0.8}Mn_{0.2}$.

**Materials and Methods**

**Chemicals and Materials**

Six distinct alloys, namely $Ni_{0.8}X_{0.2}$ and $Ni_{0.6}Co_{0.2}X_{0.2}$, where X represents Co, Fe, and Mn, were meticulously prepared. Metal pieces of Ni, Co, Fe, and Mn, each with a purity of at least 99.99 at.%, were precisely weighed and combined through arc-melting. To ensure a uniformly blended composition, the arc-melted buttons underwent at least five cycles of flipping and remelting before being drop-cast into a copper mold. The resulting as-cast ingots were homogenized at 1200 °C for 24 hours, followed by rapid water quenching. These homogenized ingots were then cut into 2 mm thick square discs using electro-discharge machining. Subsequently, the discs underwent electro-polishing to eliminate any stressed layers.

**Computational Methods**

First-principles density functional theory (DFT) calculations were conducted using the Projector Augmented Wave (PAW) method and the Perdew-Burke-Ernzerhof (PBE) exchange potential. These calculations were implemented within the Vienna Ab Initio Simulation Package (VASP) code (*30, 31*). The exploration of diffusion barriers and pathways utilized the Climbing-Image Nudged Elastic Band (CI-NEB) method (*32*), employing three intermediate images for optimization along the reaction path. Vacancy diffusion energy ($E_d$) was determined as the difference between the DFT-calculated energies of the saddle point ($E_s$) and the initial point ($E_i$) along the diffusion path, expressed as $E_d = E_s - E_i$. The tetrahedron smearing method, complemented by Blöchl corrections, was implemented in fixed dimension/volume calculations to generate the density of states (DOS). Brillouin zones were sampled using K-point grids, maintaining a uniform spacing of $2\pi \times 0.04$ Å$^{-1}$. Model structures underwent thorough optimization, with energy and force thresholds set at $10^{-4}$ eV and 0.02 eV/Å, respectively. Electron wavefunctions were expanded using a plane wave basis up to 400 eV. These parameters were meticulously chosen through pre-calculation to ensure precision in results while minimizing computational resource utilization.

The structures were modeled using the Similar Local Approximation Environment (SAE) method (*33*), which creates analogous local atomic environments for all lattice sites. It is vital to differentiate the SAE method from the Special Quasirandom Structures (SQS) method, where a cluster function utilized to establish the site occupation correlation function. In contrast, the SAE method employs a similarity function to achieve the correlation function. Studies have demonstrated that the optimal solution obtained through the SAE method effectively addresses the issue of the local lattice distortion problem and short-range ordering effect. A comprehensive comparison between SAE and SQS has been elaborated upon in our recent publication (*34*). In this study, the $Ni_{0.6}Co_{0.2}Co_{0.2}$, $Ni_{0.6}Co_{0.2}Fe_{0.2}$, and $Ni_{0.6}Co_{0.2}Mn_{0.2}$ configurations are represented by 2×3×1 supercells, each comprising a total of 120 atoms. BCC supercells are adopted for V, $V_{0.8}Mn_{0.2}$, Cr, $Cr_{0.8}Mn_{0.2}$, Fe, $Fe_{0.8}Mn_{0.2}$, Al, $Al_{0.8}Mn_{0.2}$, Ta, $Ta_{0.8}Mn_{0.2}$, W, and $W_{0.8}Mn_{0.2}$, collectively accounting for a total atomic count of 90. Additionally, FCC supercells, encompassing Cu, $Cu_{0.8}Mn_{0.2}$, Zn, and $Zn_{0.8}Mn_{0.2}$, are employed, each comprising 120 atoms. Ti and $Ti_{0.8}Mn_{0.2}$ are represented by HCP supercells, with an atomic count of 90. For all the systems, three vacancies are investigated, by randomly removing Ni atoms. The first nearest neighboring atoms are exchanged with the vacancy by the NEB method to determine the $E_d$ of the vacancy.

When evaluating transition metals, particularly Mn, a critical step involves comparing calculated lattice constants with experimental values. This assessment is essential in determining the necessity of introducing the Hubbard parameter U, which addresses the impact of strong electron-electron interactions. In equiatomic NiCoMn alloy, the calculated lattice constant (3.58 Å) closely corresponds to the experimentally determined value (3.60 Å) *(22)*. This close match obviates the need for incorporating the Hubbard parameter U in Ni-based CSAs containing Mn. However, in this study, only in the FeMn system, the lattice constant without U measures at 3.26 Å, a noteworthy deviation from the experimental value of 3.57 Å *(35)*. This prompted a meticulous calibration and validation process to determine the optimal value of U, ultimately achieving the most consistent alignment with the experimental value when U is set to 2.

**Supplementary Text**

**Factors those disregarded in classical Friedel Oscillations formula**

1) Atomic size mismatch

**Table S1** displays the atomic radius of elements in the alloy systems (*36*). The variation in atomic radius, spanning from the minimum in Ni to the maximum in Mn, is only 0.03 Å. This slight difference implies that the distortion potential energy resulting from atomic size mismatches can be considered negligible.

2) Point charge approximation

The DOS integral serves as a quantitative measure to assess the number of electrons occupancy in energy states around the resulting vacancy (*14*):

$$\Delta\rho = \Delta N(E) = N_P(E) - V_V(E) \tag{S1}$$

Where $N_P(E)$ and $N_V(E)$ represent the sum of DOS within the 1nn atomic range of vacancy before and after vacancy formation. Employing **Eq. (S1)** and taking the local atomic environment depicted in **Fig. 1A** as an illustrative example, the charge density change within the 1nn regions ($\Delta\rho_{1nn}$) in local integrated DOS below the Fermi level after Ni-vacancy formation in the CSAs are investigated, as shown in **Fig. S2** During this process, the vacancy is treated as point charges. Nevertheless, it is crucial to acknowledge that this hypothetical charge is not strictly localized to a singular point at the vacancy center; rather, it may be distributed across its spatial extent. Utilizing the charge density matrix from the DFT output file CHGCAR, we computed the charge density within the vacancy region, yielding values of 0.17, 0.18, and 0.25 for $Ni_{0.6}Co_{0.2}Co_{0.2}$, $Ni_{0.6}Co_{0.2}Fe_{0.2}$, and $Ni_{0.6}Co_{0.2}Mn_{0.2}$ alloys, respectively. These values represent, at most, 3% of $\Delta\rho_{1nn}$, emphasizing the minimal impact of the point charge approximation on the FO formula.

3) Scattering phase shift

Rennert (*37*) et al. proposed the incorporation of the scattering intensity characterized by Lorentzian properties:

$$t(\omega) = \frac{1}{\pi\rho_0} \frac{\Delta}{\omega - \omega_0 + i\Delta} \tag{S2}$$

At the resonance energy $\omega_0$, $\rho_0$ represents the unperturbed conduction-electron density of states for one spin direction, and $\Delta$ denotes the width of the resonance, the function $t(\omega)$ exhibits correlations with both the energy transfer's proximity to the Fermi level and the width of the scattering process. In accordance with the free-electron Green's function (*38*), the FO formula can be refined as follows:

$$\Delta\rho'(r) = \int_{-E_f}^{\infty} (2l+1)\pi^{-1} \text{Im}\left\{t_l(\omega)\left[G_l(r;\omega+i\delta)\right]^2\right\} f(\omega)d\omega \tag{S3}$$

The Fermi function, denoted as $f(\omega)$, and the *l*th angular-momentum component ($G_l$) of the free-electron Green's function are involved in the expression. Utilizing the Riemann lemma and the dispersion relation approximation (*18*), the outcomes can be represented in a corrected Friedel formula:

$$\Delta\rho'(r) = -\left[\frac{(2l+1)}{4\pi^2}\right]\sin\delta(0)\left[\frac{a(r)}{r^3}\right]\cos\left[2k_F - l\pi + \delta_l(0) + 2\varphi(k_F r) - \eta(r)\right] \quad \text{(S4)}$$

By disregarding the phase shift ($\pi\rho_0 t_l(0) = -\sin\delta_l(0)\exp[i\delta_l(0)]$), **Eq. (S4)** aligns with the classical Friedel formula, as expressed in **Eq. (3)**. While considering the phase shift *(39)*, **Eq. (S4)** can be evolved into:

$$\Delta\rho'(r) = \frac{A\cos(2k_F + \varphi)}{r^3} + \frac{B\sin(k'_F + \varphi)}{r^2} \quad \text{(S5)}$$

In which $A$ and $B$ are the amplitude. $\varphi$ is the phase of the charge oscillations.

Note that, in this study, it is essential to take account into the stochastic dispersion of Mn atoms, wherein distinct quantities of Mn atoms are situated on surfaces at varying distances along the direction from vacancy. The different number of Mn atoms per atomic plane involved will cause different scattering superposition effects. Accordingly, a variable 'n,' representing the number of Mn atoms, should be incorporated into **Eq. (S5)** to precisely account for the superposition effect of Mn. $K_F$ is the Fermi level of the system, which is obtained through DFT calculation with the value of 4.8 eV. $k'_F$ is the Fermi level of Mn atom, which is 5.3 eV. The value of $A$, $B$ and $\varphi$ is 1.4, 0.5, and -5, respectively, in order to match with the lattice constant. As such, the perturbation charge $\Delta\rho'(r)$ along <001> direction in $Ni_{0.6}Co_{0.2}Mn_{0.2}$ caused by V1 vacancy shown in **Fig. 2C** can be refitted by considering the phase shift effects:

$$\Delta\rho'(r) = \frac{1.4\times\cos(5.8r - 5)}{r^3} - \frac{0.5n\times\sin(9.6r - 5)}{r^2} \quad \text{(S6)}$$

The results obtained from fitting **Eq. (S6)** show excellent agreement with the actual process, confirming that the scattering phase shift generated by Mn is the primary factor causing anomalous FO.

**Balancing intra-to-conduction-electron and intra-electron interactions**

1) Electron-electron interaction from Slater theory

In accordance with the Anderson Hamiltonian (16), the Hamiltonian $H$ is defined as the sum of four terms, denoted as $H_1$, $H_2$, $H_3$, and $H_4$, expressed as:

$$H = H_1 + H_2 + H_3 + H_4$$
$$= \sum_{k,\sigma} \eta_{k\sigma} c'_{k\sigma} c_{k\sigma} + \sum_{m,\sigma} \varepsilon_d d'_{m\sigma} d_{m\sigma} + \sum_{k,m,\sigma} V_{km}(d'_m c_{k\sigma} + c'_{k\sigma} d_m) + \sum_{ijkl} U d'_i d_j d'_k d_l$$

Here, $H_1$ represents the unperturbed energy of the free-electron system in second-quantized notation. $H_2$ corresponds to the unperturbed energy of the "$d$" states on the atom. $H_3$ is the interaction term between $d$ electrons and conduction electrons. The fourth term, $H_4$, accounts for the repulsive energy among the $d$ functions and is given by:

$$H_4 = \frac{1}{2}(U_{eff} - J)\sum_{m \neq n} d'_{m\sigma} d_{m\sigma} d'_{n\sigma} d_{n\sigma} + \frac{1}{2} U_{eff} \sum_{m,n} d'_{m\sigma} d_{m\sigma} d'_{n\sigma} d_{n\sigma} \tag{S7}$$

The intra-elemental electrons interaction energy $U_{eff}$ in **Eq. (S7)** is expressed in terms of two Slater integrals, the monopole contribution to the Coulomb energy $F^0$ and the multipole integral $F^2 + F^4$. According to the Slater theory, exemplified by $U_{eff}(d^5)$, which can be expressed as:

$$U_{eff}(d^5) = \left[E(d^6(^5D)) - E(d^6(^6D))\right] - \left[E(d^4(^5D)) - E(d^5(^6S))\right] = F^0 + \frac{4}{14}\left(F^2 + F^4\right)$$

The free atom exhibits a Coulomb interaction among intra-elemental electrons, as their spins align according to Hund's rule. However, in a solid, this coupling is partially disrupted by interactions with neighboring atoms. In our scenario, $H_3$ signifies the hybridization between vacancies and Mn atoms, while $H_4$ represents the effective Coulomb interactions among Mn intra-elemental $3d$ electrons, which are directly influenced by $U_{eff}$. The interplay between $H_3$ and $H_4$ profoundly impacts the electron arrangement.

2) Experimental determination of $U_{eff}$

In this study, $U_{eff}$ is determined through Auger Electron Spectroscopy (AES), X-ray photoelectron spectroscopy (XPS), and ultraviolet photo-electron spectroscopy (UPS) experiments. X-ray photoelectron and x-ray excited Auger spectra were acquired using a Thermo ESCALAB 250XI spectrometer with an Al $K_\alpha$ source and a linewidth of 0.1 eV. The UPS measurements were conducted using the Thermofisher ESCALAB 250Xi instrument. Prior to measurement, the samples underwent argon-ion etching and in situ heat treatment. The light source utilized was a He resonance lamp emitting photons at 21.22 eV. For the cases presented in this study, a 10 eV pass energy was utilized. In $3d$ transition metals, electronic transition process LMM is one process that directly exhibits the interaction between two $3d$ electrons (16). The initial step in Auger spectroscopy involves ionizing a $2p_{3/2}$ electron, leading to the ion's decay to a low-energy state through the Auger process. The $2p$ hole is replenished by one $3d$ electron, and another $3d$ electron is ejected, resulting in the final state with two holes in the $3d$ shell. Consequently, the kinetic energy of the ejected Auger electron is contingent on the energy of the $2p$ hole ($E_{2p}$), $3d$ hole ($E_{3d}$), and $U_{eff}$. We can then write for the Auger process with two holes in the $3d$ shell:

$$E_k = E_{2p} - 2E_{3d} - U_{eff} \tag{S8}$$

where $E_{2p}$ and $E_{3d}$ are the energy required for electrons to be excited and $E_k$ is the kinetic energy of the Auger electron. The $E_{2p}$ and $E_{3d}$ can be measured by XPS and UPS. **Fig. S5** show the 2p$_{3/2}$ lines XPS spectra and Auger spectrum of Ni, Co, Fe, and Mn elements for the Ni$_{0.6}$Co$_{0.2}$X$_{0.2}$ alloys in this study. **Fig. S6** illustrates UPS spectra of Ni$_{0.8}$Co$_{0.2}$, Ni$_{0.6}$Co$_{0.2}$Co$_{0.2}$, Ni$_{0.6}$Co$_{0.2}$Fe$_{0.2}$, and Ni$_{0.6}$Co$_{0.2}$Mn$_{0.2}$ alloys. The minimum energy required to remove a 3d electron above the Fermi level is calculated using: $E_{3d}$ = 21.22 eV - $E_0$, where $E_0$ represents the cutoff energy. In our investigation, $E_0$ for Ni$_{0.8}$Co$_{0.2}$, Ni$_{0.6}$Co$_{0.2}$Co$_{0.2}$, Ni$_{0.6}$Co$_{0.2}$Fe$_{0.2}$, and Ni$_{0.6}$Co$_{0.2}$Mn$_{0.2}$ is measured as 2.87 eV, 3.08 eV, 3 eV, and 3.51 eV, respectively. **Table S2** outlines the experimental findings for both binding energies and kinetic energies. The $U_{eff}$ values determined in our study for Ni, Co, and Fe are consistent with those documented in previous publications (*16, 39-42*), affirming the reliability of our experimental approach. It is obvious that the $U_{eff}$ of Mn is notably larger than that of Ni, Co, and Fe.

3) Coulomb attraction induced by vacancy

Following the creation of a vacancy, it is considered to carry a positive charge resulting from the reduction of electrons. This positive charge induces an attractive force on the conduction electrons of the neighboring atoms, a force comprehensible through the Coulomb relationship. In accordance with this relationship, the Coulombic effect of a vacancy on an electron can be articulated as follows:

$$E_c = \frac{kq_1q_2}{r} = \frac{9\times10^9 \times 0.25\times1.6\times10^9 \times 1\times1.6\times10^9}{2.8\times10^{-10}} \approx 2 \ eV \tag{S9}$$

The Coulomb attraction induced by vacancy generation is markedly smaller than the restored $U_{eff}$ for Mn (~4.6 eV).

4) Crystal Orbital Hamilton Populations (COHP) calculations

In the Ni$_{0.6}$Co$_{0.2}$X$_{0.2}$ (X=Co/Fe/Mn) alloy system, depicted in **Fig. S7**, we employ the Crystal Orbital Hamilton Population (COHP) method to explore the electron hybridization between X atoms and Ni atoms creating vacancies, with particular attention to the V1 case outlined in **Fig. 1**. Our analysis unveils that in the case of Mn, nearly two electrons are engaged in hybridization, a substantially higher figure compared to Fe and Co, where participation is less than one electron. This heightened hybridization effect in Mn can be attributed to its unique half-filled 3d electronic structure. The abundance of single empty orbitals in Mn leads to a notably broadened density of states, fostering robust hybridization with neighboring atoms in alloys (*18*).

Post-vacancy formation, the cessation of hybridization triggers the restoration of correlation effects among Mn electrons. As mentioned above, due to the large $U_{eff}$ of Mn, the electron compensation towards the vacancy direction is depleted. According to the **Eq. (2)**, the decrease in potential energy caused by vacancy weakens the energy at the saddle point position on the diffusion path, ultimately promoting the accelerated diffusion of vacancies.

**Diffusion couple experiments and resistance measurements**

1) Diffusion couple experiments

To demonstrate the specific influence of Mn on atomic diffusion, we carefully designed the diffusion couple experiments. Three diffusion couples $Ni_{0.8}X_{0.2}/Ni_{0.6}Co_{0.2}X_{0.2}$ (X=Co/Fe/Mn), were prepared, with the diffusion couple fixture, as depicted in **Fig. S8**, serving as a crucial component for the experiment. The fixture material, a Kovar alloy (Fe-29Ni-17Co nickel–cobalt ferrous alloy), was chosen for its similar thermal expansion coefficient to the Ni-based alloys used in the study. Axial force was applied using a tension wrench with a tightening torque of 15 N·m for each of the diffusion couples. Tantalum sheets were introduced between the hex cap screw and the diffusion couples to ensure uniform stress distribution across the samples. The loaded fixture was encased in a quartz capsule, which underwent evacuation and subsequent refilling with inert helium gas to 0.03 MPa for all diffusion couples. This measure was specifically taken to prevent oxidation and volatilization of Mn, particularly in the $Ni_{0.8}Mn_{0.2}/Ni_{0.6}Co_{0.2}Mn_{0.2}$ case.

Each type of diffusion couple was prepared in triplicate, and these three sets of couples were annealed at temperatures of 900 °C, 1000 °C, and 1100 °C for 100 hours, respectively. After annealing, the diffusion couples were cross-sectional cut, ground and chemical-mechanical polished. The concentration profiles were measured using Electron Probe Micro-Analyzer (EPMA) with electron energy at about 15 keV. The cross-sectional morphology diagram and the composition profile of three diffusion couples after annealed at 900 °C, 1000 °C, and 1100 °C for 100 hours is shown in **Fig. S9**. For each diffusion couple, three line-scans were conducted across the cross-section to measure the concentration profiles accurately. The lateral resolution of our measurement is around 1 μm and all data points in **Fig. 4A** are directly from measurement counts without smoothing or fitting. Thus, the scattering of these points represents the uncertainties of the composition measurement.

By examining the curve depicting the variation in element concentration, we can determine the diffusion coefficients ($D$) for both Ni and Co.

According to the Fick's second law:

$$(\frac{\partial c}{\partial t})_x = \left\{\frac{\partial}{\partial x}\left(D\frac{\partial C}{\partial x}\right)\right\}_t \tag{S10}$$

The initial and boundary conditions for the concentration at position $x$ and time $t$, $c(x, t)$, are

$$\begin{cases} C(\eta=0)=C_1 \\ C(\eta=\infty)=C_0 \end{cases} \tag{S11}$$

In which $\eta = x/\sqrt{t}$ is the Boltzmann variable. By choosing the Matano interface to ensure an equilibrium diffusion flow density on both sides and designating the corresponding coordinate $X_M$ as the coordinate origin, as depicted in **Fig, S10**, the $D$ can be determined as:

$$D = -\frac{1}{2t(\frac{\partial C}{\partial x})_{x=x_0}} \int_{c_0}^{c} x_M dc \tag{S12}$$

In **Fig, S10**, the fitted Boltzmann distribution curve for Co element in $Ni_{0.8}Co_{0.2}$/$Ni_{0.6}Co_{0.2}Co_{0.2}$ couple after annealing at 1100 °C for 100 hours is:

$$y = A_1 + \frac{A_2}{1+\frac{x-x_0}{dx}} = 39.84 + \frac{-19.63}{1+\frac{x+15.9}{19.48}} \quad (S13)$$

Solving **Eq. S13** yields the Co diffusion coefficient with the value of $1.447\times10^{-15} m^2/s$. Similarly, the $D$ of Co in the other two diffusion couples can be determined by adopting the Boltzmann-Matano method as mentioned above, as shown in **Table S3**. At equivalent temperature, the $D$ of Co in diffusion couple containing Mn is significantly higher than that in diffusion couples without Mn.

At a given temperature, the magnitude of $D$ is governed by the diffusion activation energy ($Q$). In our case, the $D$ obtained for Co from the diffusion concentration profiles reflects the $Q$ for Co in the $Ni_{0.6}Co_{0.2}X_{0.2}$ phase. This is because the diffusion of Co primarily occurs in the direction from $Ni_{0.6}Co_{0.2}X_{0.2}$ to $Ni_{0.8}X_{0.2}$, and the concentration profiles will predominantly represent the behavior of Co in the $Ni_{0.6}Co_{0.2}X_{0.2}$ region.

The Arrhenius equation establishes the connection between diffusion coefficient $D$ and diffusion activation energy $Q$. The logarithm of $D$ is graphed as a linear function of l/T, unveiling the diffusion activation energy ($Q$) through the slope. The value of $Q$ for Co atoms in the samples are listed in the **Table S4**. In $Ni_{0.6}Co_{0.2}Mn_{0.2}$, the value of Q is 2.61 eV, which is lower than that in $Ni_{0.6}Co_{0.2}Co_{0.2}$ with 3.08 eV and in $Ni_{0.6}Co_{0.2}Fe_{0.2}$ with 3.22 eV.

2) Electrical resistivity measurements

The $Ni_{0.6}Co_{0.2}X_{0.2}$ (X= Co/Fe/Mn) and $Ni_{0.6}X_{0.2}$ (X= Co/Fe/Mn) samples were cut into 2×3×12 mm cuboid using electro-discharge machining. Subsequently, the cuboid underwent electro-polishing to eliminate any stressed layers before being used in the resistance measurement experiment. The sample's resistance measurement is conducted in two temperature ranges: low-temperature section (2-300 K) and high-temperature section (300-800 K). For low-temperatures, a 4-wire technique is employed with an AC current (0.2 mA, 128 Hz) in the Physical Property Measurement System (Quantum Design). Golden wires (30 microns diameter) bonded to the samples with silver paste facilitate this measurement. The high-temperature electrical resistance is determined using the LSR-2/800 Seebeck coefficient/electric resistance measuring system from LINSEIS Germany.

$E_f$ can be determined through electrical resistivity measurements at various temperatures. According to Mott and Jones theory *(26, 27)*, the electrical resistance is anticipated to show a quadratic dependence at low temperatures:

$$R = A + BT + CT^2 \quad (S14)$$

However, as the temperature increases, the vacancy formations lead to an increase in the concentration of vacancy thermal equilibrium, resulting in a higher measured actual resistance than the value calculated by **Eq. (S14)**. The resistance at high temperature $R_h$ can be expressed as:

$$R_h = A + BT + CT^2 + A^{'}\exp(E_f/kT) \quad (S15)$$

In which $E_f$ is the vacancy formation energy. At the same temperature, the difference between $R_h$ and the extrapolated ones from low temperature measured values, by using **Eq. (S14)**, $\Delta R$, has the relationship with the vacancy formation energy as:

$$\Delta R = A' \exp(E_f / kT) \quad (S16)$$

By solving **Eq. (S16)**, the vacancy formation energy of the corresponding system can be obtained. For $Ni_{0.6}Co_{0.2}Co_{0.2}$, fit the resistance measured under 300 K using the quadratic function as:

$$y = 6.97 \times 10^{-13} x^2 + 1.2 \times 10^{-10} x + 2.9 \times 10^{-8} \quad (S17)$$

The extrapolation of **Eq. (S17)** to elevated temperatures (300-800 K) facilitated the determination of the deviation $\Delta R$, obtained by subtracting the extrapolated values from the experimental results at the same temperature. Subsequently, a linear correlation is established between the natural logarithm of the change in resistance (ln$\Delta R$) and the reciprocal of temperature (1/T), as illustrated in **Fig. S11.** This methodology is applied to other samples as well. The slope of the resulting linear correlation is indicative of the vacancy formation energy $E_f$, measured at 1.63 eV for $Ni_{0.6}Co_{0.2}Co_{0.2}$, 1.70 eV for $Ni_{0.6}Co_{0.2}Fe_{0.2}$, and 1.66 eV for $Ni_{0.6}Co_{0.2}Mn_{0.2}$, showing no significant differences. It is noteworthy that the vacancy equilibrium concentration is similar on both sides of the three diffusion couples, indicating that the key factor influencing differences in macro-level vacancy diffusion lies primarily in diffusion barriers $E_d$. The experimental obtained $E_d$ are listed in **Table S5**. These values are very consistent with our DFT calculation results, which strongly prove that the influence of Mn on the microscopic diffusion barrier of vacancies can be extended to the macroscopic diffusion performance.

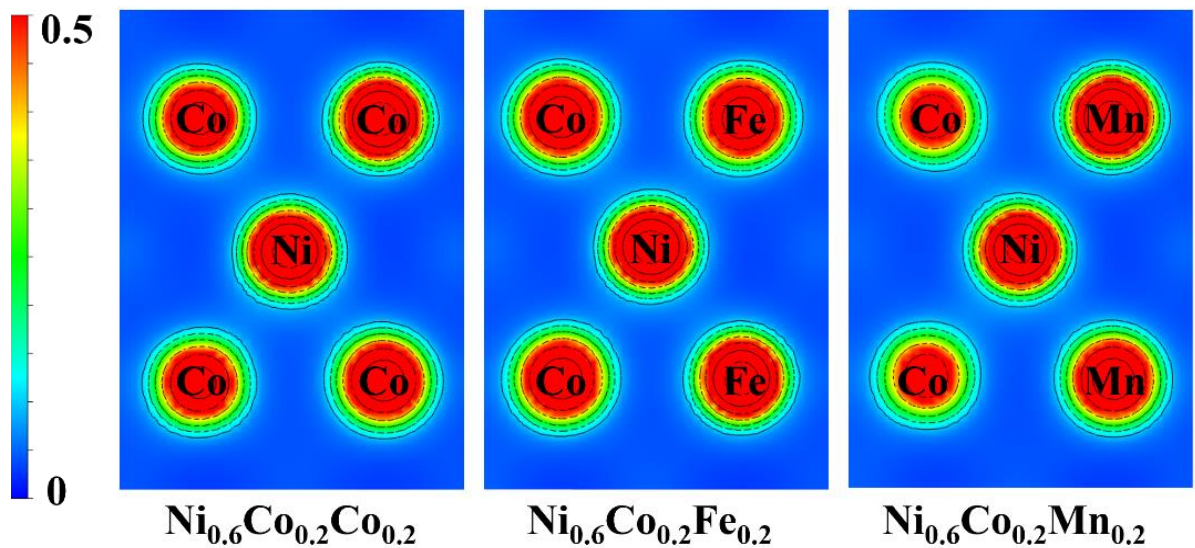

**Fig. S1.** Two-dimensional charge density distribution of the Ni diffusing atoms and their 1nn atoms at the saddle points (SP) in $Ni_{0.6}Co_{0.2}Co_{0.2}$, $Ni_{0.6}Co_{0.2}Fe_{0.2}$, and $Ni_{0.6}Co_{0.2}Mn_{0.2}$ alloys. Notably, even at the SP, the closest position to the 1nn atoms, the diffusing atom experiences minimal interactions with them.

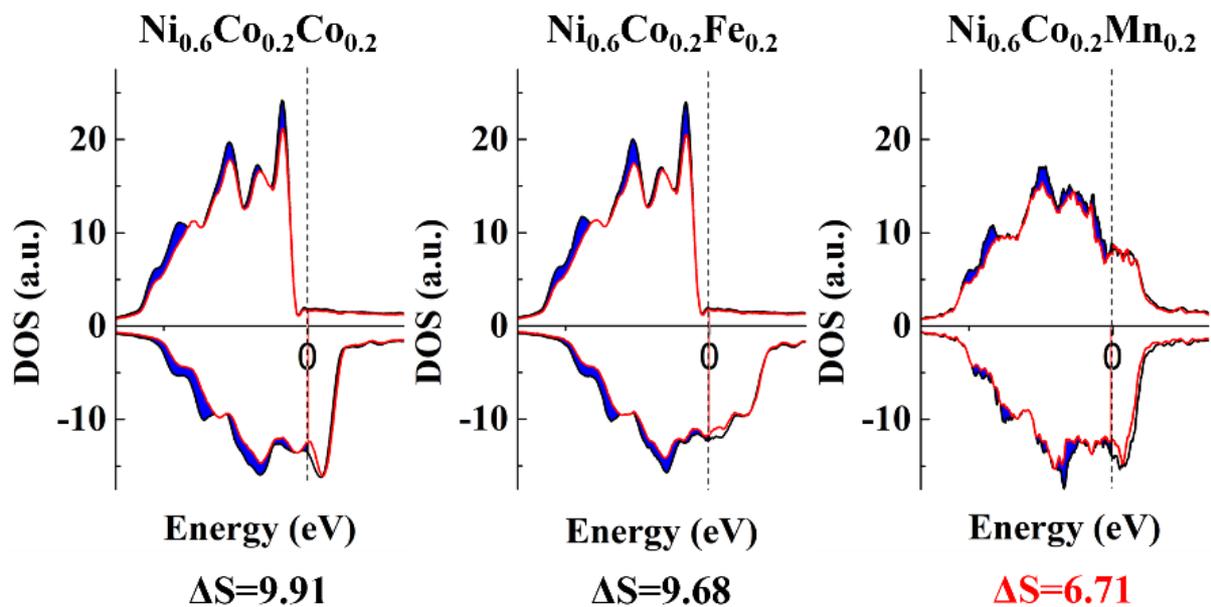

**Fig. S2.** The amount of charge density changes within the 1nn regions ($\Delta\rho_{1nn}$) of vacancy (shaded in blue) before (black line) and after (red curve) the vacancy formation. The area of the blue region ΔS represents $\Delta\rho_{1nn}$. In $Ni_{0.6}Co_{0.2}Mn_{0.2}$, $\Delta\rho_{1nn}$=6.71.

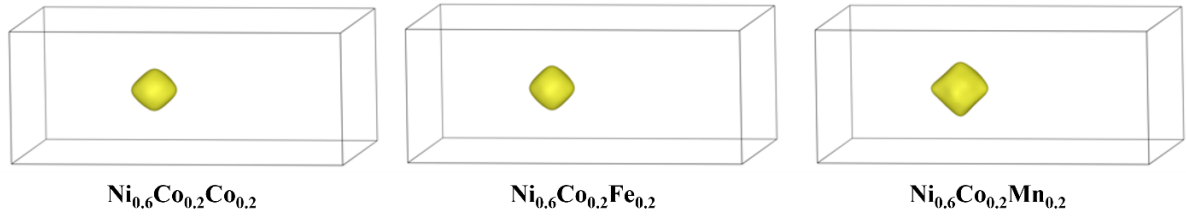

**Fig. S3.** The charge density within the vacancy region after vacancy formation using the same isosurface of 0.015 e/Å, yielding values of 0.17, 0.18, and 0.25 for $Ni_{0.6}Co_{0.2}Co_{0.2}$, $Ni_{0.6}Co_{0.2}Fe_{0.2}$, and $Ni_{0.6}Co_{0.2}Mn_{0.2}$ alloys, respectively. These values represent, at most, 3% of $\Delta\rho_{1nn}$, emphasizing the minimal impact of the point charge approximation on the FO formula.

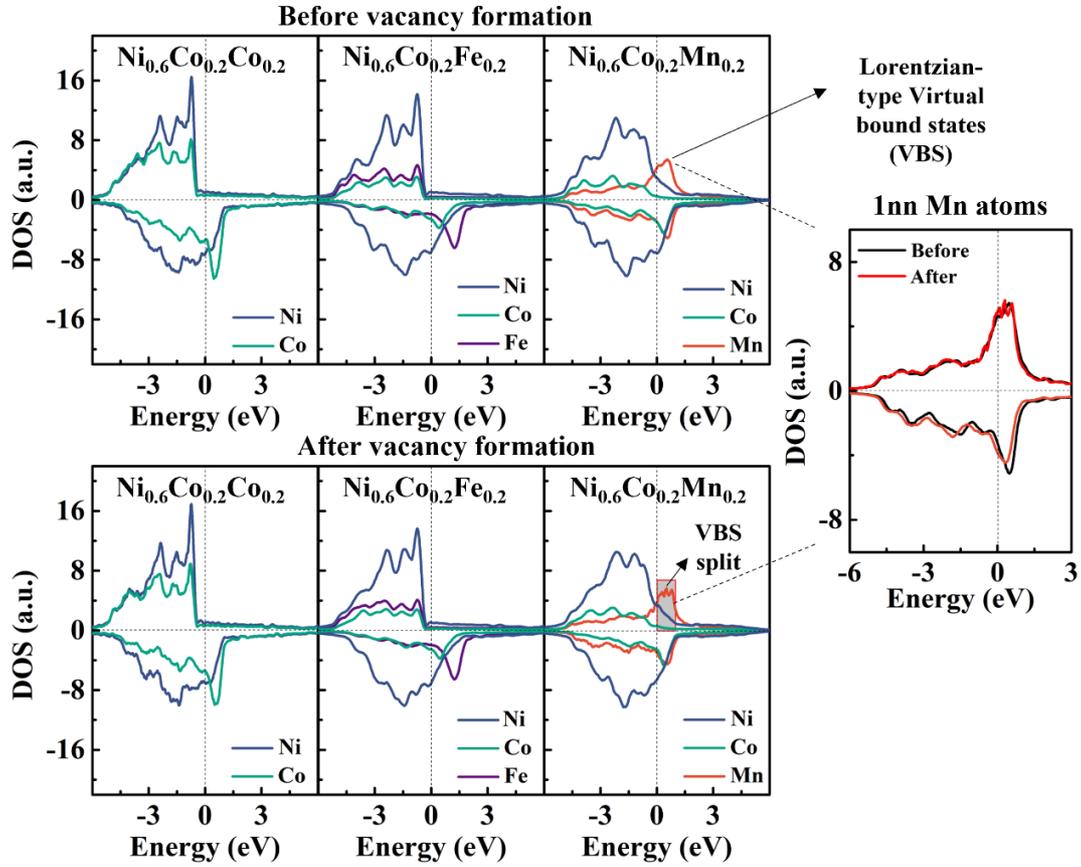

**Fig. S4.** The PDOS of the 1nn atoms before and after vacancy formation. Pre-vacancy, the Mn DOS exhibits Lorentzian-type virtual bound states (VBS) above the host *d* bands. Post-vacancy formation, the splitting of the VBS in Mn suggest a local moment self-consistent renormalization.

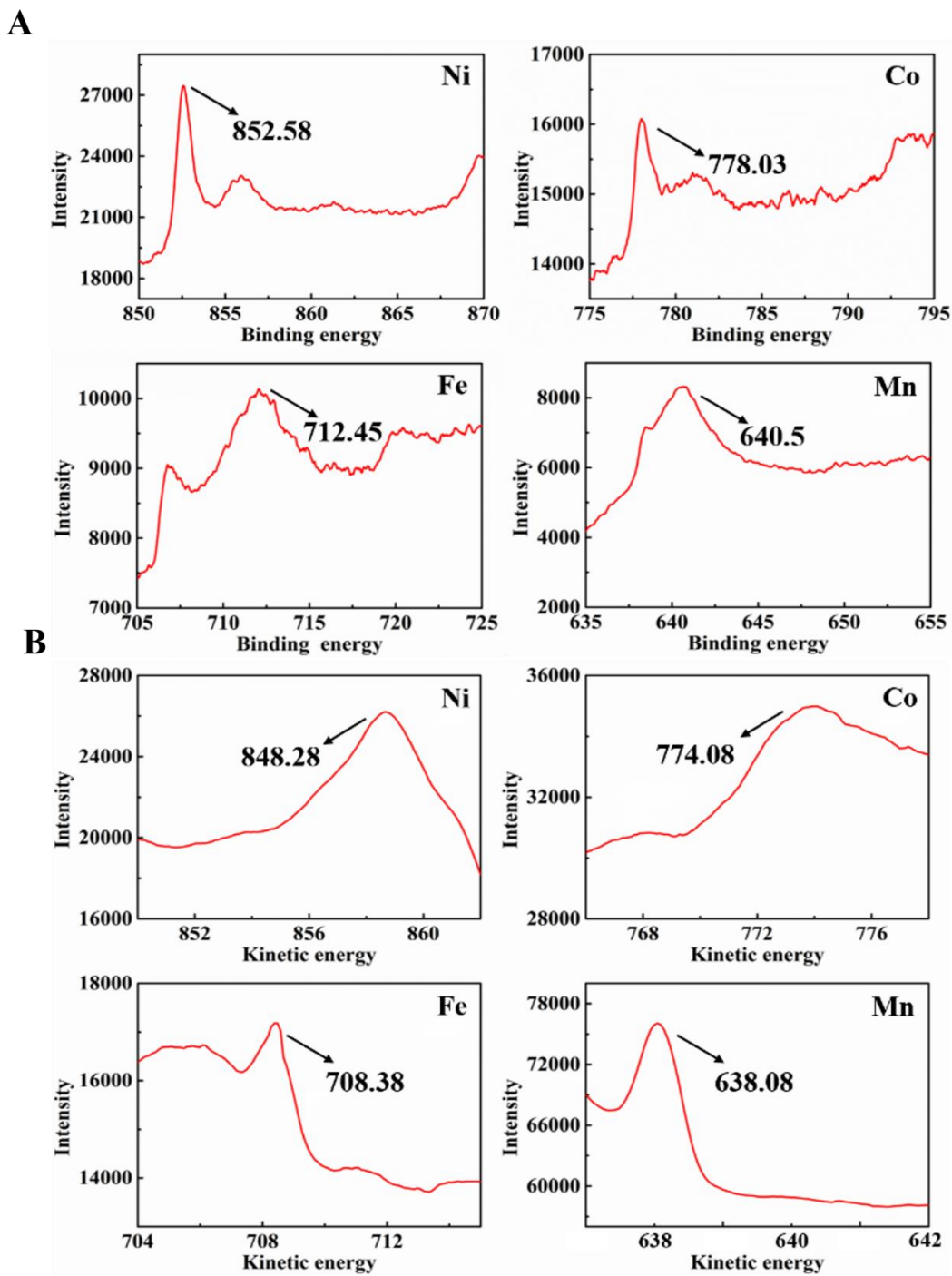

**Fig. S5.** $2p_{3/2}$ lines XPS spectra (**A**) and AES spectrum (**B**) of the Ni, Co, Fe, and Mn for the $Ni_{0.6}Co_{0.2}X_{0.2}$ alloys in this study. In the LMM transition process, the intra-elemental electrons interaction energy $U_{eff}$ can be calculated by the kinetic energy of the ejected Auger electron and the energy of electrons. The detailed calculation process is in the "experimental determination of $U_{eff}$" section on page 6 in this supplementary material.

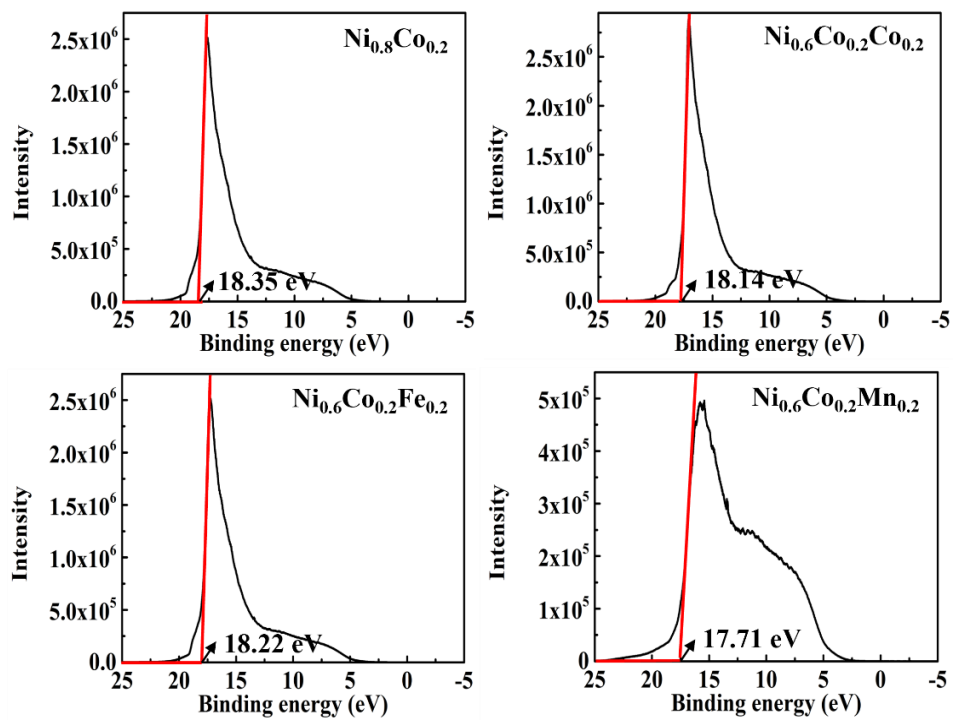

**Fig. S6.** UPS spectra of $Ni_{0.8}Co_{0.2}$, $Ni_{0.6}Co_{0.2}Co_{0.2}$, $Ni_{0.6}Co_{0.2}Fe_{0.2}$, and $Ni_{0.6}Co_{0.2}Mn_{0.2}$ alloys.

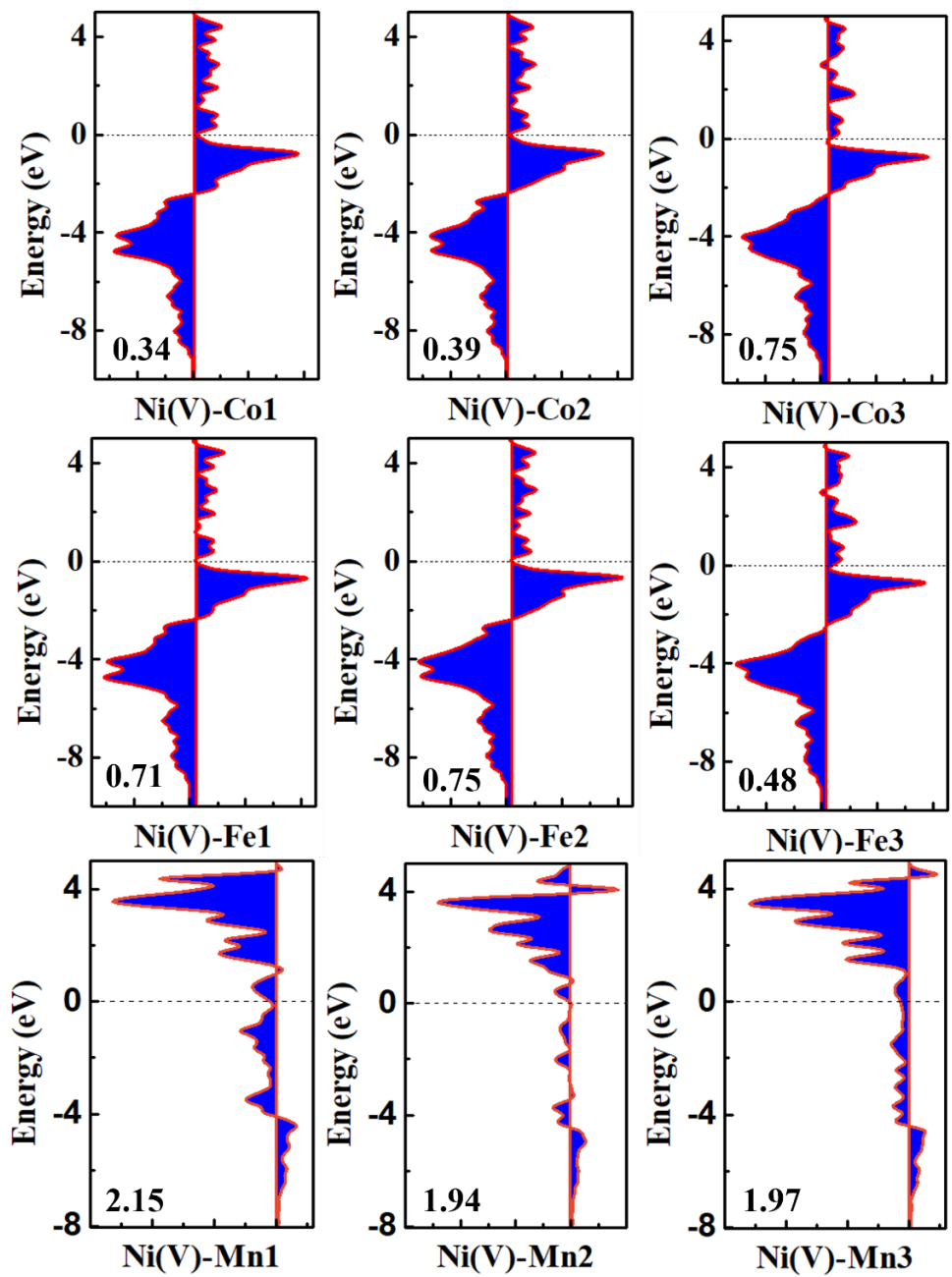

**Fig. S7**. The COHP curves of hybridization between Ni atom that generated vacancy Ni(V) and its 1nn atoms in $Ni_{0.6}Co_{0.2}X_{0.2}$ (X=Co/Fe/Mn). The Fermi levels, $E_F$, are represented by the black horizontal dash lines. In the case of Mn, nearly two electrons are engaged in hybridization, a substantially higher value compared to Fe and Co, where participation is less than one electron.

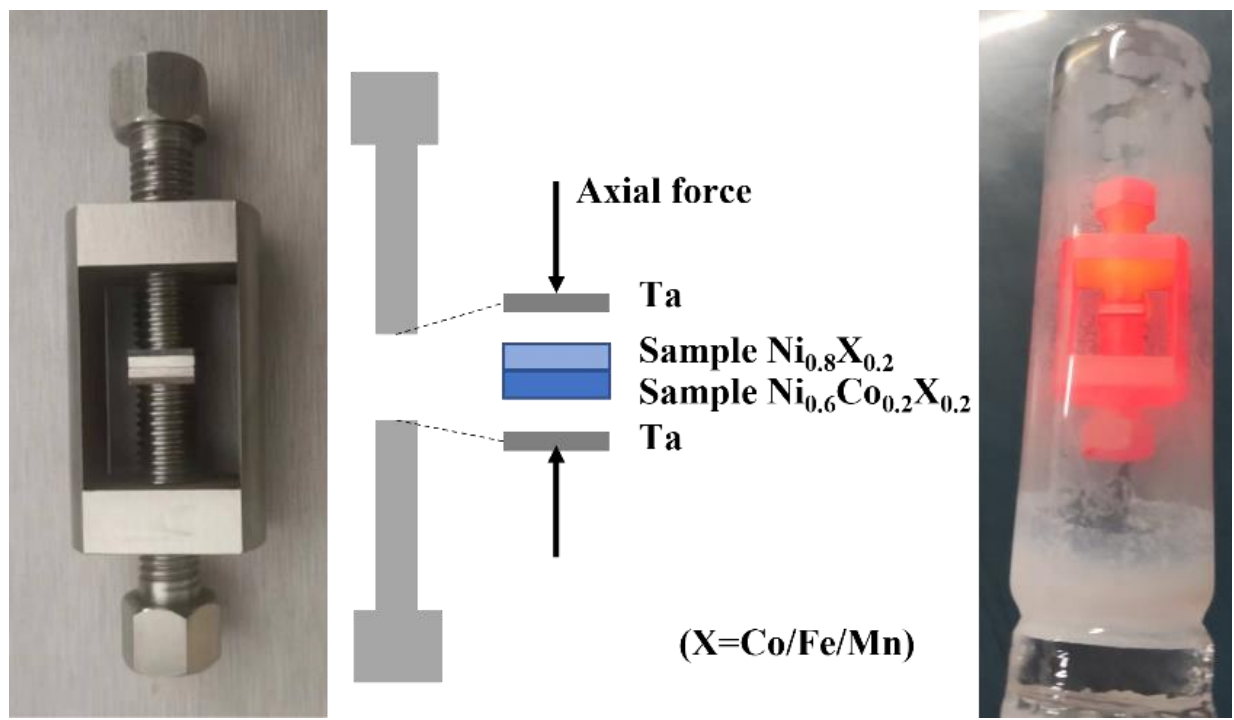

**Fig. S8.** Diffusion couple assembly made by Kovar alloy and the schematics for experimental.

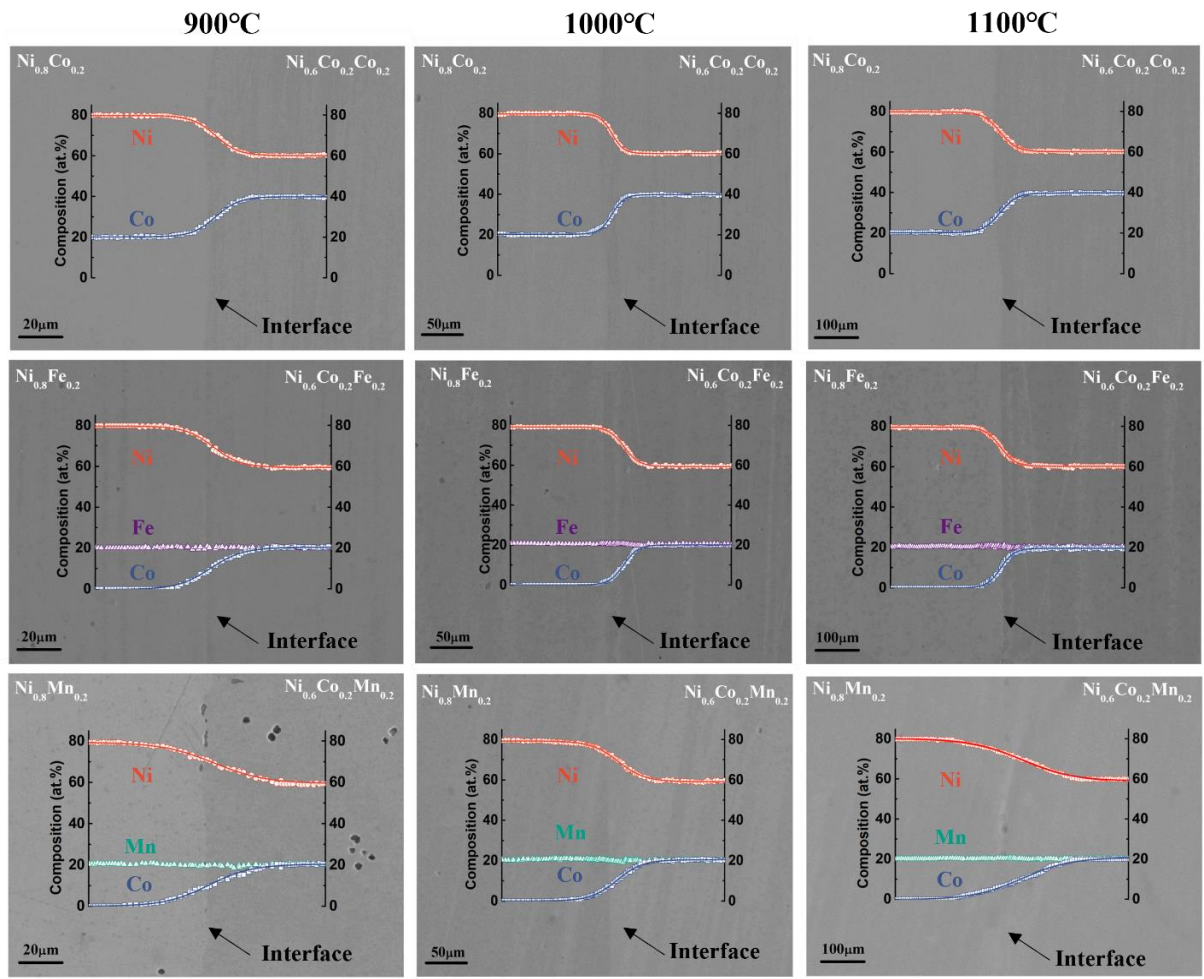

**Fig. S9.** The cross-sectional morphology diagram measured by EPMA and the composition profile of the three diffusion couples annealed at 900 °C, 1000 °C, and 1100 °C for 100 hours.

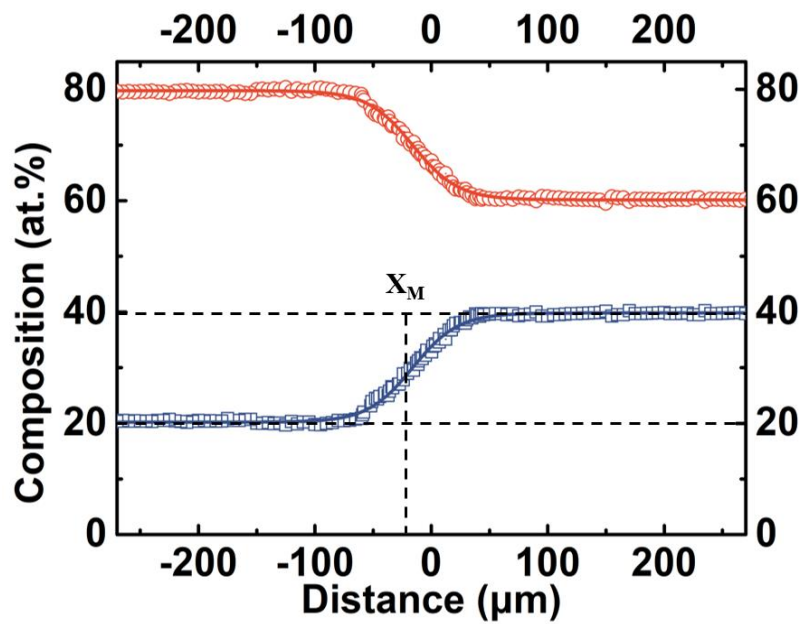

**Fig. S10.** The diffusion profile and Matano interface for Co in $Ni_{0.8}Co_{0.2}$/ $Ni_{0.6}Co_{0.2}Co_{0.2}$ couple after annealing at 1100 °C for 100 hours. The detailed calculation process is shown in the "diffusion couple experiments" section on page 8 in this supplementary material.

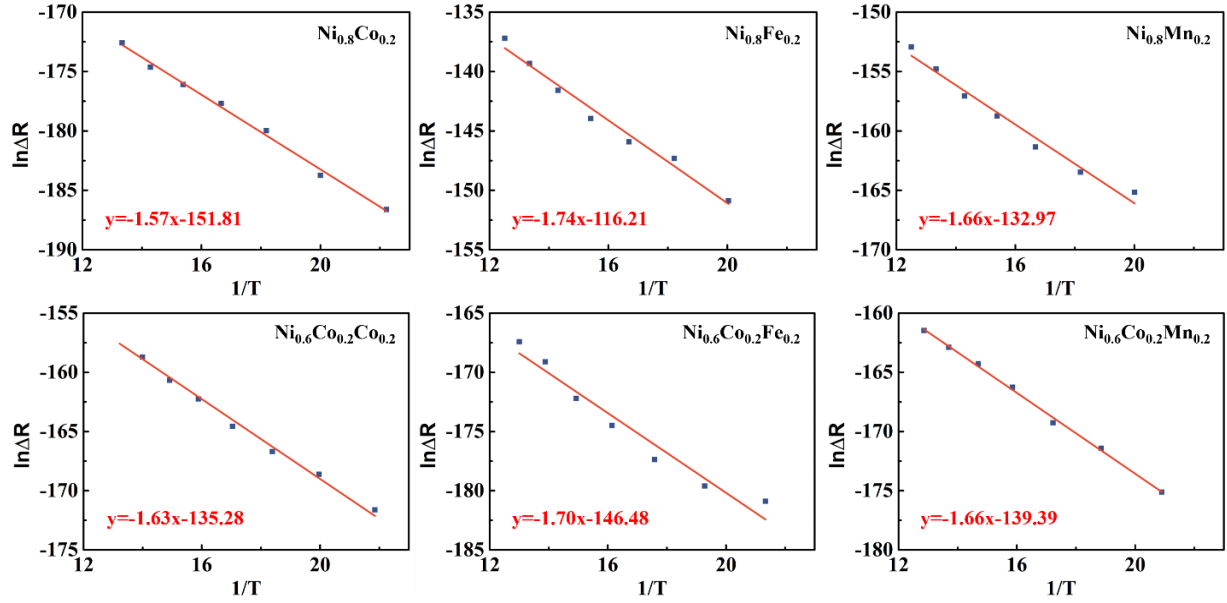

**Fig. S11.** The correlation between the natural logarithm of $\Delta R$ (ln$\Delta R$) and the inverse of temperature (1/T). $\Delta R$ is the deviation between the resistance value obtained from the extrapolation function and the resistance value measured experimentally at the same temperature. The extrapolation function is a quadratic function fitted by the resistance at low temperatures (under 300 K) and extrapolate to elevated temperatures (300-800 K). The gradient of this relationship signifies the vacancy formation energy $E_f$. The $E_f$ for all samples are ranged from 1.57 eV to 1.74 eV. It illustrates that the vacancy equilibrium concentration is similar on both sides of the three diffusion couples, indicating that the key factor influencing differences in macro-level vacancy diffusion lies primarily in diffusion barriers $E_d$. The calculation formula is detailed in the "electrical resistivity measurements" section on page 9 in this supplementary material.

**Table S1.** The atomic radius of the elements in the alloy systems (*34*). The variation in atomic radius, spanning from the minimum in Ni to the maximum in Mn, is only 0.03 Å. This slight difference implies that the distortion potential energy resulting from atomic size mismatches can be considered negligible.

|                  | Ni   | Co   | Fe   | Mn   |
|------------------|------|------|------|------|
| Atomic radius (Å) | 1.24 | 1.25 | 1.26 | 1.27 |

**Table S2.** Corresponding experimental electronic energy parameters of atoms in the Ni$_{0.6}$Co$_{0.2}$X$_{0.2}$ alloys: core-level binding energies $E_{2p}$ obtained by XPS; the Auger kinetic energy $E_k$; core-level binding energies $E_{3d}$ obtained by UPS; effective Coulomb energy $U_{eff}$ in the LMM process; and the $U_{eff}$ from reference (*16,42*). All values are in eV. Mn has the largest value of $U_{eff}$ in these elements with 4.61 eV, which is notably larger than that of Ni, Co, and Fe. The $U_{eff}$ values determined in our study for Ni, Co, and Fe are consistent with those documented in previous publications, affirming the reliability of our experimental approach.

|  | Ni | Co | Fe | Mn |
| --- | --- | --- | --- | --- |
| $E_{2p}$ | 852.58 | 778.03 | 712.45 | 640.5 |
| $E_k$ | 848.28 | 774.08 | 708.38 | 638.08 |
| $E_{3d}$ | 2.87 | 3.08 | 3 | 3.51 |
| $|U_{eff}|$ | 1.44 | 2.21 | 1.93 | 4.61 |
| $U_{eff}$[a] | 1.61 | 2.22 | 1.83 | / |

(a) Reference 16, 39-42.

**Table S3.** The diffusion coefficients (*D*) for Co element in three diffusion couples at various temperatures. The diffusion coefficient in Mn-containing diffusion couple is two orders of magnitude higher than those in the other two.

| $D$ ($10^{-17}$m$^2$/s) | $Ni_{0.8}Co_{0.2}$/ $Ni_{0.6}Co_{0.2}Co_{0.2}$ | $Ni_{0.8}Fe_{0.2}$/ $Ni_{0.6}Co_{0.2}Fe_{0.2}$ | $Ni_{0.8}Mn_{0.2}$/ $Ni_{0.6}Co_{0.2}Mn_{0.2}$ |
|---|---|---|---|
| 1173K | 23.5 | 39.2 | 1085 |
| 1273K | 60.1 | 94.8 | 2579 |
| 1373K | 144.7 | 248.2 | 4811 |

**Table S4.** The value of Q for Co atoms in $Ni_{0.6}Co_{0.2}Co_{0.2}$, $Ni_{0.6}Co_{0.2}Fe_{0.2}$, and $Ni_{0.6}Co_{0.2}Mn_{0.2}$.

|  | $Ni_{0.6}Co_{0.2}Co_{0.2}$ | $Ni_{0.6}Co_{0.2}Fe_{0.2}$ | $Ni_{0.6}Co_{0.2}Mn_{0.2}$ |
| --- | --- | --- | --- |
| Q (eV) | 3.08 | 3.22 | 2.61 |

**Table S5**. The diffusion energy ($E_d$) obtained from the experiments and compared with the DFT calculation results ($E_{d\text{-}DFT}$), demonstrating a high level of consistency. It strongly proves that the influence of Mn on the microscopic diffusion barrier of vacancies can be extended to the macroscopic diffusion performance. All values are in eV.

|  | $E_d$ by experiment | $E_{d\text{-}DFT}$ by DFT |
|---|---|---|
| $Ni_{0.6}Co_{0.2}Co_{0.2}$ | 1.45 | 1.18 |
| $Ni_{0.6}Co_{0.2}Fe_{0.2}$ | 1.52 | 1.14 |
| $Ni_{0.6}Co_{0.2}Mn_{0.2}$ | 0.95 | 0.81 |